\documentclass[aps,prfluids,preprint,superscriptaddress,nofootinbib,longbibliography]{revtex4-2}

\usepackage{amsmath,amssymb,amsfonts,amsthm}
\usepackage{graphicx}
\usepackage{bm}
\usepackage{xcolor}
\usepackage{float}
\usepackage{physics}
\usepackage{my_macros}
\usepackage[colorlinks=true,allcolors=blue]{hyperref}

\newcommand{\Js}{J_s}
\newcommand{\KA}{K_A}

\newcommand{\rhost}{\rho^{\ast}}
\newcommand{\grads}{\nabla_s}
\newcommand{\divs}{\nabla_s\!\cdot}
\newcommand{\dbar}[1]{\overline{\overline{#1}}}

\begin{document}

\title{Stability and equilibria of a compressible elastic membrane in Stokes flow}

\author{Sho Kawakami}
\affiliation{Department of Mathematical Sciences, New Jersey Institute of Technology, Newark, New Jersey 07102, USA}

\author{Han Zhou}
\affiliation{Department of Mathematics, University of Pennsylvania, Philadelphia, Pennsylvania 19104, USA}

\author{Po-Chun Kuo}
\affiliation{Department of Mathematics, Purdue University, West Lafayette, Indiana 47907, USA}

\author{Yoichiro Mori}
\email{y1mori@sas.upenn.edu}
\affiliation{Department of Mathematics, University of Pennsylvania, Philadelphia, Pennsylvania 19104, USA}

\author{Yuan-Nan Young}
\email{yyoung@njit.edu}
\affiliation{Department of Mathematical Sciences, New Jersey Institute of Technology, Newark, New Jersey 07102, USA}

\date{\today}

\begin{abstract}
% We determine the dispersion relation for a fluid bilayer membrane, taking into
% account the coupling between hydrodynamic stress, elastic bending and the local variation of lipid density due to the finite area modulus that renders the bilayer membrane weakly inextensible. Expressing the membrane tension as an explicit function of lipid density difference, we show that, in the inextensible limit, our model yields the same shape bifurcations as those in the well-studied inextensible membrane model where the membrane tension is a Lagrange multiplier. Furthermore, our results make direct connections between membrane tension and lipid density difference between the two monolayers, allowing more explicit mathematical analysis and physical interpretation of membrane shape and local variation of lipid density difference. 
We formulate a continuum model for a compressible lipid-bilayer membrane immersed in Stokes flow, replacing exact local area inextensibility by conservation of an areal phospholipid density. The membrane free energy combines Helfrich bending, spontaneous curvature, and a finite area-compression penalty, so that membrane tension becomes a constitutive response to lipid-density variation rather than a Lagrange multiplier enforcing local area conservation. The resulting interfacial stress includes normal elastic forces and tangential Marangoni stresses generated by lipid redistribution; these stresses arise from membrane compressibility and can produce an effective negative tension when the local lipid density exceeds its preferred value.
We further derive the linear stability of circular membranes in two dimensions and spherical membranes in three dimensions under full Stokes hydrodynamic coupling. In both cases, bending stabilizes the base shape, while excess lipid density destabilizes it by favoring increased membrane area. The first instability occurs in the lowest nontrivial shape mode, \(m=2\) in two dimensions and \(j=2\) in three dimensions. Energy expansions near onset show that the two-dimensional instability is a pitchfork bifurcation, whereas the three-dimensional instability is generically transcritical because prolate and oblate perturbations are geometrically distinct. These results provide a controlled compressible extension of classical vesicle mechanics and directly connect lipid-density variation, membrane tension, hydrodynamic coupling, and shape instability.
\end{abstract}


\maketitle

\section{Introduction}

Vesicles, sacks of viscous fluid enclosed by lipid bilayer membranes, have been extensively studied by theory, simulations,  and experiments for understanding the features of cell mechanics that arise from the fluid-structure interactions of a fluid-phase lipid bilayer membrane.  At nanoscopic scales, a lipid bilayer is not merely a material surface, but a thin three-dimensional structure whose thickness participates directly in its mechanics, transport, and coupling to the surrounding fluid. This distinction is often hidden in classical membrane theories, where the bilayer is reduced to a two-dimensional interface and described through Helfrich-type energetics. Such surface-based models have been remarkably successful in connecting membrane mechanics to phenomena such as endocytosis, adhesion, excitability, protein diffusion, and large-scale remodeling \cite{iwasa1980mechanical,watanabe2013ultrafast,shi2014dynamics,carlson2015protein,ling2020high,helfrich1973elastic,steigmann1999fluid,deserno2015fluid,sahu2017irreversible,liu2006endocytic,agrawal2009modeling,dmitrieff2015membrane,omar2020nonaxisymmetric,saffman1975brownian,agrawal2011model,samanta2021vortex}. Yet several experimentally relevant effects arise precisely from the finite separation between the two membrane faces: hydrophobic mismatch couples protein function to local thickness \cite{phillips2009emerging}, short-wavelength fluctuations retain signatures of thickness degrees of freedom \cite{woodka2012lipid}, and permeability to small solutes varies systematically with bilayer thickness \cite{frallicciardi2022membrane}. Phenomenological extensions of Helfrich theory have incorporated some of these corrections and have provided insight into relaxation dynamics, density fluctuations, and structure-factor measurements in membrane–fluid systems \cite{seifert1993viscous,evans1994hidden,merkel1989molecular,yeung1995unexpected,watson2011intermediate,fournier2015hydrodynamics,terzi2017novel,terzi2019consistent,levine2014determination,hamm2000elastic,pinigin2020additional,deseri2008derivation,faizi2024curvature,rahimi2012shape,kelley2023nanoscale,nagao2017probing}. \textcolor{black}{These descriptions provide important reduced models, but they also motivate the question of how finite-thickness and density effects should be represented within an interfacial continuum theory coupled to the surrounding fluid.}

A central simplification in classical vesicle mechanics is the assumption that the lipid
bilayer is locally inextensible. This assumption is physically motivated by the large area
expansion modulus of lipid membranes, typically
$K_A^{\rm phys}\sim 0.2\text{--}0.3\,{\rm N/m}$, with micropipette measurements reporting values
near $243\,{\rm mN/m}$ for several phosphatidylcholine bilayers. Thus, over the
low-tension regime relevant to many vesicle experiments, changes in apparent membrane
area arise primarily from the smoothing of thermal undulations rather than from direct
molecular dilation of the bilayer. Once the thermal excess area is exhausted, further
stretching is penalized strongly, with the true areal strain scaling as
\begin{equation}
\frac{\Delta A}{A}\sim \frac{\gamma}{K_A^{\rm phys}}.
\end{equation}
For tension $\gamma$ of order $0.1\text{--}0.5\,{\rm mN/m}$, this estimate gives molecular
areal strains of order $10^{-3}$, or roughly $0.04\%\text{--}0.2\%$. In this sense,
lipid bilayers are not exactly inextensible materials; rather, their large stretching
modulus makes local inextensibility an accurate continuum approximation whenever true
areal dilation is small compared with bending, hydrodynamic, and geometric effects.

Standard vesicle theories therefore impose local area conservation, either in single-layer
membrane models~\cite{vlahovska2007dynamics} or in double-layer descriptions~\cite{shi2014dynamics}. Mathematically,
this constraint is enforced by an in-plane tension field that acts as a Lagrange multiplier,
ensuring that the surface velocity remains locally area-preserving. This formulation is
appropriate when lipid number conservation can be identified directly with area conservation.
In the present work, we relax this identification by introducing a membrane model with explicit lipid-density variation, thereby capturing density changes that may occur under flow in strong confinement \cite{peng2026fluid}. Instead of imposing local area
inextensibility as a constraint, we formulate lipid conservation through an areal number
density defined on the membrane surface. Local changes in membrane area are then allowed,
but are penalized by a density-dependent areal free energy. The resulting membrane tension
is no longer a Lagrange multiplier; it is a constitutive response generated by deviations of
the lipid density from its reference value. In this way, the present formulation provides a
controlled compressible extension of the standard continuum theory for fluid-phase vesicle
membranes.

\textcolor{black}{Formulated with an areal lipid density, the model shows how lipid redistribution generates Marangoni stresses through the constitutive tension field. We use the term Marangoni stress in this mechanical sense: the tangential traction has the form $-\nabla_\Gamma\gamma$, but the tension gradients are generated by density conservation and membrane compressibility rather than by surfactant adsorption or desorption. These stresses can produce an effective negative tension when the local lipid density exceeds its preferred value.}

\textcolor{black}{The main contribution of this work is to connect this constitutive-tension mechanism to analytically tractable shape instabilities. A conserved lipid-density field replaces the local inextensibility constraint, the resulting density-dependent tension couples lipid redistribution to Stokes flow, and excess density destabilizes the base circular or spherical shape by favoring increased membrane area. The same linear spectrum also reveals a fast density-dominated relaxation mode, which provides a mechanics-induced density-relaxation scale. Near onset, the first unstable shape is the lowest nontrivial mode, but the bifurcation structure depends on dimension: the two-dimensional circular problem gives a pitchfork bifurcation, whereas the three-dimensional spherical problem is generically transcritical because prolate and oblate perturbations are not geometrically equivalent.}

\textcolor{black}{We use this framework to study instability induced by membrane area-density mismatch in both two and three dimensions. In the classical vesicle literature, area mismatch often appears through reduced volume, bilayer-couple constraints, or area-difference elasticity. Here, those nonlocal or global descriptors are replaced by a local density mismatch field, allowing us to examine how density-dependent tension modifies linear thresholds and onset branches.}
 
\textcolor{black}{This paper is organized as follows. Section~\ref{sec:model} presents the model formulation for compressible membranes immersed in Stokes flow. Section~\ref{sec:spectrum} gives the linear stability analysis for a circular membrane in two dimensions and a spherical membrane in three dimensions. Section~\ref{sec:variational} analyzes the bifurcation of membrane shape with respect to the area modulus and compares the two- and three-dimensional settings. Section~\ref{sec:conclusion} summarizes the results and discusses future directions.}

\section{Model}\label{sec:model}
\textcolor{black}{In this section, we introduce a free energy depending on membrane shape and phospholipid density, and derive the dynamical equations for the membrane coupled to Stokes flow. We use this model to examine the stability of circular and spherical base shapes for a closed extensible membrane in two and three dimensions in Section~\ref{sec:spectrum}.}

\subsection{Free Energy}
\begin{figure}
    \begin{center}\includegraphics[width=1.0\linewidth]{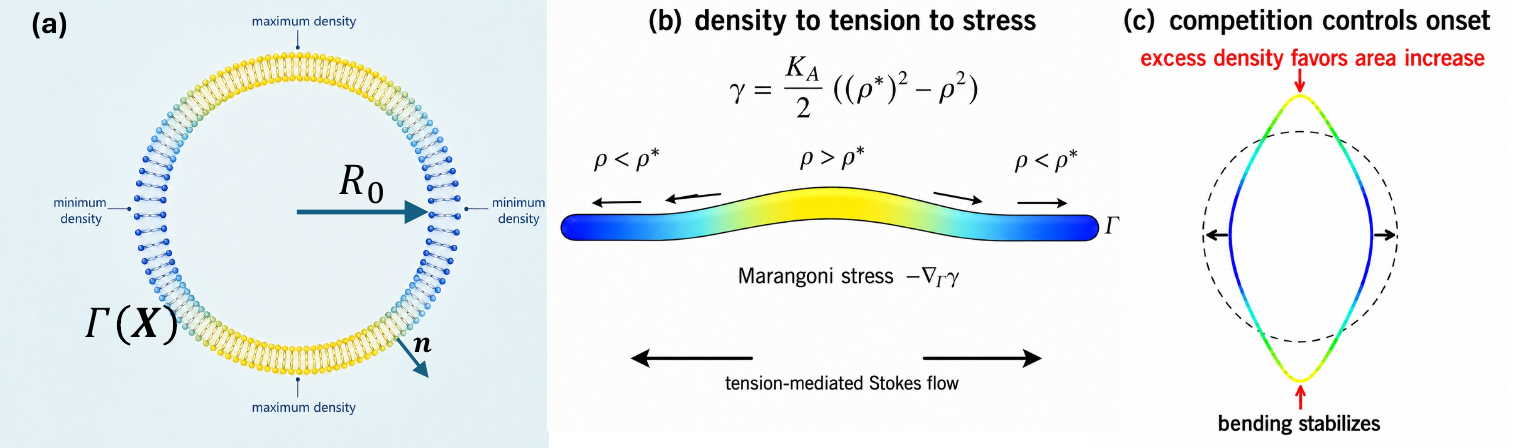}\caption{\textcolor{black}{Schematic of the compressible membrane model. (a)\&(b): A spatially varying lipid density $\rho$ generates a constitutive tension field $\gamma(\rho)$; surface gradients of this tension produce Marangoni stresses that couple lipid redistribution to Stokes flow and membrane shape dynamics. (c):  The equilibrium membrane shape is determined by the  balance between stabilizing bending forces and density-driven elastic tension.}}\label{fig:illustration}
    \end{center}
\end{figure}

% \textcolor{black}{Consider a closed membrane with enclosed fluid volume $V$ in three dimensions and enclosed area $A$ in two dimensions, with boundary $\surf$ and outward normal $\bn$.} The total lipid mass in the membrane is fixed at $4\pi\rhobar$.
% The free energy of a surface compressible lipid bilayer membrane is given by
% \begin{equation}
%     E = \int_\surf \frac{\Bendmod}{2}(\meancurv-\meancurv_0)^2+\Bendmod_G\gausscurv+\frac{\Areamod^{\rm phys}}{2}(\rho-\rho^\ast)^2 \,d\surf.\label{eq:energydim}
% \end{equation}
% $\meancurv=-\frac{1}{2}\bnab_\surf\cdot\bn$ is the local mean curvature of the interface where $\bnab_\surf = (\bI-\bn\bn)\cdot\bnab$. This definition of the mean curvature may different from the definition in other works, such as \cite{zhong1989bending}, by a factor of $-1/2$. $\gausscurv$ is the Gaussian curvature, $\Bendmod\in[0,\infty)$ is the bending modulus, $\meancurv_0\in\R$ is the spontaneous curvature of the interface.
% $\rho\in[0,\infty)$ is the local phospholipid density, hereby simply referred to as lipid, density, with $\rho^\ast\in[0,\infty)$ the preferred density. 
% $\Areamod\in[0,\infty)$ is the area compression modulus.
%\col{Note to self: Need to check $\Bendmod\to\Bendmod/2$ for all equations in 3d}
\textcolor{black}{Consider a closed membrane with enclosed fluid volume $V$ in three dimensions and enclosed area $A$ in two dimensions, with boundary $\surf$ and outward normal $\bn$, see \reffig{fig:illustration}(a).} 
The free energy of a surface compressible phospholipid bilayer membrane is
\begin{equation}
    E = \int_\surf \frac{\Bendmod}{2}(\meancurv-\meancurv_0)^2+\Bendmod_G\gausscurv+\frac{\Areamod^{\rm phys}}{2}\left(\frac{\rho}{\rho^\ast}-1\right)^2 \,d\surf,\label{eq:energydim}
\end{equation}
similar to the free energy of a membrane with lipid density inhomogeneities\cite{bitbol2011membrane}.
\textcolor{black}{The first two terms in the integrand of \refeqn{eq:energydim} are the Helfrich bending energy \cite{canham1970minimum,helfrich1973elastic,evans1974bending}. The local mean curvature is defined by $\meancurv=-\frac{1}{2}\bnab_\surf\cdot\bn$, where $\bnab_\surf=(\bI-\bn\bn)\cdot\bnab$ is the surface gradient operator. This convention differs from some other definitions, such as that in \cite{zhong1989bending}, by a factor of $-1/2$. Here $\gausscurv$ is the Gaussian curvature, $\Bendmod\in[0,\infty)$ is the bending modulus, and $\meancurv_0\in\R$ is the spontaneous curvature. The third term penalizes deviations of the local phospholipid density $\rho$ from the preferred density $\rho^\ast$, with dimensional penalty coefficient $\Areamod^{\rm phys}\in[0,\infty)$.}

% The first and second term in the integrand is associated to the Helfrich bending energy \cite{}, which is minimized by the spherical/circular membrane shape. 
% The integral of the second term containing the Gaussian curvature is constant for closed membranes independent of shape and will be omitted from the analysis as it does not affect energy minimization or membrane dynamics calculations.
% The third term corresponds to energy due to deviation of local density $\rho$ from a preferred density $\rho^\ast$.
% The average local density is dependent on the surface area of the interface.
% In this study $\rho^\ast <\rhobar$, the energy of the density term is therefore minimized by diluting local density by increasing the surface area of the membrane.
% Under the constraint of fixed volume, this leads to competing effects, with the bending energy stabilizing the spherical/circular shape and the density component driving instability of the spherical/circular shape.
% Further detail on the derivation of the free energy due to density variation can be found in \refapp{appsec:FreeEnergy}.
The Helfrich bending energy is minimized by the spherical/circular membrane shape. 
The integral of the second term containing the Gaussian curvature is constant for closed membranes independent of its shape and will be omitted from further analysis because it does not affect energy minimization or membrane dynamics calculations.
The third term corresponds to energy due to deviation of local density $\rho$ from the preferred density $\rho^\ast$, see \reffig{fig:illustration}(b).
In the regime considered here, \(\rho^\ast < \rhobar\), an average density for a reference shape $\Gamma_0$ (\(\rhobar\equiv\int_{\Gamma_0}\rho d\Gamma_0/\int_{\Gamma_0}d\Gamma_0\)). The density-dependent contribution to the membrane energy is therefore reduced by lowering the local lipid density, which, for fixed total lipid mass, is achieved by increasing the membrane surface area. Under the constraint of fixed enclosed volume, this tendency competes with bending elasticity: the bending energy favors the spherical/circular configuration, whereas the density energy favors area-increasing deformations. The resulting competition can destabilize the spherical/circular shape, illustrated in \reffig{fig:illustration}(c).
Further detail on the derivation of the free energy due to density variation can be found in \refapp{appsec:FreeEnergy}. 

\subsection{Governing equations for flows}
Assuming the Reynolds number is sufficiently small, the flow velocity $\bu$ and the pressure $\press$ satisfy the Stokes equation,
\begin{equation}
    -\bnab\press^{(i)}+\visc\bnab^2\bu^{(i)} = \bzero,\quad \bnab\cdot\bu^{(i)} = 0 \label{eq:3DStokeseq}
\end{equation}
where $(i)\in \{\ins,\out\}$ indicates inside and outside the drop respectively.
At the interface $\surf$, the velocity is continuous
\begin{equation}
    \bu_\surf := \bu^\ins = \bu^\out,\;\; \text{ for } \bx\in\surf.\label{eq:3DBC1}
\end{equation}
The flow vanishes far away from the membrane
\begin{equation}
    \bu^\out \to0, \;\; \text{ as } |\bx|\to\infty.\label{eq:3DBC2}
\end{equation}
Finally there is the stress balance equation
\begin{equation}
    [[\bm{\sigma}\cdot\bn]] = \tau^B+\tau^E,\label{eq:3DBC3}
\end{equation}
\textcolor{black}{where $[[\bm{\sigma}\cdot\bn]]=(\bm{\sigma}^\out-\bm{\sigma}^\ins)\cdot\bn$ is the jump in the bulk hydrodynamic stress $\bm{\sigma}^{(i)}=-\press^{(i)}\bI+\visc(\bnab\bu^{(i)}+(\bnab\bu^{(i)})^T)$.}
The stress is derived from the energy as
\begin{equation}
    \tau = \frac{\partial }{\partial \delta}E(\bX+\delta\bX)|_{\delta=0}\label{eq:energystress}
\end{equation}
where $\delta\bX$ is an arbitrary perturbation of membrane shape $\bX$.
The first term on the right hand side of \refeqn{eq:3DBC3} corresponds to the bending stresses
\begin{equation}
    \tau^B = \Bendmod(\bnab_\surf^2\meancurv+2(\meancurv-\meancurv_0)(\meancurv^2+\meancurv\meancurv_0-\gausscurv))\bn.\label{eq:bendingstress}
\end{equation}
%where $\bnab_\surf = (\bI-\bn\bn)\cdot \bnab$ is the surface gradient operator and $H = -\frac{1}{2}\bnab_\surf \cdot\bn$ is the mean curvature.
The second term on the right hand side of \refeqn{eq:3DBC3} corresponds to an elastic term (the membrane elasticity):
\begin{equation}
\tau^E = -(2\meancurv\ten\bn+\bnab_\surf\ten),\quad \ten = \frac{K_A^{\rm phys}}{2}\left[1-\left(\frac{\rho}{\rho^\ast}\right)^2\right].\label{eq:elasticstress}
\end{equation}
When the local density is near the preferred density, the tension can be linearized about $\rho=\rho^\ast$:
\begin{equation}
    \ten = K_A^{\rm phys} \left(1-\frac{\rho}{\rho^\ast}\right).\label{eq:elasticstress2}
\end{equation}
This linearization of the density-tension relation and the resulting tension gradient is similar to those observed in models for surfactant-covered drops. 
The distinguishing feature from  surfactant models is the negative tension that is observed when $\rho>\rho^\ast$. 
This negative tension from the energetic view is a reduction in the free energy due to dilution of lipid density from increased surface area. 
From a dynamic point of view, the negative tension works to grow perturbations normal to the interface and destabilize it if the local density is too large.
\textcolor{black}{For a material point $\bx_\surf\in\surf$, the interface is advected by the interfacial velocity,} \begin{equation}
    \frac{d\bx_\surf}{dt} = \bu_\surf,
    \label{eq:3Dshapeevo}
\end{equation}
and the evolution of the lipid density is given by 
\begin{equation}
    \frac{\partial\rho}{\partial t} +\bnab_\surf\cdot(\rho\bu_\surf) = 0,\label{eq:3Ddensevo}
\end{equation} 
as the transport of the local stretching of the interface in \cite{aland2014diffuse}.
An illustration of the problem and a summary of the dynamics can be found in \reffig{fig:illustration}.

{\color{black}
Although \refeqn{eq:3Ddensevo} contains no explicit Fickian surface-diffusion term, this does not imply that density perturbations are passively frozen into the membrane. In the present model, density variations generate gradients of the constitutive tension, and these Marangoni stresses drive Stokes flows that relax the density field. The linear stability analysis in Section~\ref{sec:spectrum} shows that this tension-mediated relaxation produces a fast density-dominated mode, which may be interpreted as a mechanics-induced effective density diffusivity (see \refeqn{eq:mechD} in Section~\ref{subsec:spherical_membrane}) . For typical vesicle-scale parameters, this relaxation is much faster than molecular lipid diffusion, and we therefore omit an additional molecular diffusion term in the present formulation.
}
% In the present model, the lipid density is treated in the purely advective
% limit. This assumption may be interpreted either as a material conservation law
% written with respect to the lipid barycentric surface velocity, or as the
% large-surface-Peclet-number limit in which lateral diffusive relaxation is slow
% compared with the hydrodynamic deformation time. The relevant dimensionless
% parameter is the surface Peclet number
% \begin{equation}
%     {\mathrm Pe}_s = \frac{U_s L_s}{D_s},
% \end{equation}
% where \(U_s\) is a characteristic membrane surface velocity, \(L_s\) is the
% lateral length scale of the density variation, and \(D_s\) is the lateral lipid
% diffusivity. For a fluid lipid bilayer membrane, \({\mathrm Pe}_s\) can vary substantially
% with the deformation rate and length scale, but a biologically relevant range is
% roughly \({\mathrm Pe}_s \sim 10^{-2}\)--\(10^{2}\). Thus the purely advective model should be understood as a large-\({\mathrm Pe}_s\) approximation rather than as the statement that real lipid diffusivity is literally zero. A diffusive, or more generally
% thermodynamically driven, surface flux would be expected to affect slow dynamics
% and short-wavelength density variations.

\subsection{Rescaled energy and dynamics}\label{secsub:rescaled}
The energy is normalized with respect to the preferred density $\rho^\ast$, length scale $\Rdrop$, and bending modulus $\kappa$.
Here $\Rdrop =(3V/4\pi)^{1/3}$ in three dimensions and $\Rdrop=(A/\pi)^{1/2}$ in two dimensions.
With
\[
\meancurv=\frac{\widetilde{\meancurv}}{\Rdrop},\quad
\meancurv_0=\frac{\widetilde{\meancurv}_0}{\Rdrop},\quad
\rho=\rho^\ast\widetilde{\rho},\quad
d\surf=\Rdrop^{n-1}d\widetilde{\surf},\quad
E=\kappa\Rdrop^{n-3}\widetilde E,
\]
the nondimensional energy is
\begin{equation}
    \widetilde E=\int_{\widetilde{\surf}} \frac{1}{2}(\widetilde{\meancurv}-\widetilde{\meancurv}_0)^2+\frac{\widetilde{\Areamod}}{2}(\widetilde{\rho}-1)^2 d\widetilde{\surf}.\label{eq:energyrescaled}
\end{equation}
where $\widetilde{\Areamod}=K_A^{\rm phys}\Rdrop^{2}/\kappa$.
The nondimensional total lipid count on the spherical reference shape is $4\pi\chi$, where $\chi=\rhobar/\rho^\ast$.
We define the nondimensional area-compression modulus by $K_A\equiv\widetilde{\Areamod}$. After nondimensionalization, $K_A$ denotes this dimensionless modulus unless explicitly labeled as $K_A^{\rm phys}$.

For the dynamic equation, the viscosity is incorporated via the viscous time scale $T=\mu \Rdrop^3/\kappa$.
With
\[
t=T\widetilde t,\quad \bx=\Rdrop\widetilde{\bx},\quad
\bu=\frac{\Rdrop}{T}\widetilde{\bu},\quad
\press=\frac{\kappa}{\Rdrop^3}\widetilde{\press},\quad
\bsigma=\frac{\kappa}{\Rdrop^3}\widetilde{\bsigma},
\]
%\col{(Po: We may replace $R_0^3$ with $R_0^{6-n}$.)}
the nondimensional equations are
\begin{equation}
    \begin{split}
        &-\widetilde{\bnab}\widetilde{\press}^{(i)}+\widetilde{\bnab}^2\widetilde{\bu}^{(i)} = \bzero,\quad \widetilde{\bnab}\cdot\widetilde{\bu}^{(i)} = 0,\\
        &\widetilde{\bu}_m = \widetilde{\bu}^\ins = \widetilde{\bu}^\out \text{ for } \widetilde{\bx}\in\widetilde{\surf},\qquad
        \widetilde{\bu}^\out \to0 \text{ as } |\widetilde{\bx}|\to\infty,\\
        &[[\widetilde{\bsigma}\cdot\bn]] = \widetilde{\tau}^B+\widetilde{\tau}^E,\\
        &\widetilde{\tau}^B = \left(\widetilde{\bnab}_{\widetilde{\surf}}^2\widetilde{\meancurv}
        +2(\widetilde{\meancurv}-\widetilde{\meancurv}_0)(\widetilde{\meancurv}^2+\widetilde{\meancurv}\widetilde{\meancurv}_0-\widetilde{\gausscurv})\right)\bn,\\
        &\widetilde{\tau}^E = -(2\widetilde{\meancurv}\widetilde{\ten}\bn+\widetilde{\bnab}_{\widetilde{\surf}}\widetilde{\ten}),\quad \widetilde{\ten} = \frac{\widetilde{\Areamod}}{2}(1-\widetilde{\rho}^2),\\
        &{\color{black}\frac{d\widetilde{\bx}_{\widetilde{\surf}}}{d\widetilde t}=\widetilde{\bu}_{\widetilde{\surf}},}\qquad
        \frac{\partial\widetilde{\rho}}{\partial \widetilde t} +\widetilde{\bnab}_{\widetilde{\surf}}\cdot(\widetilde{\rho}\widetilde{\bu}_\surf) = 0.
        \end{split}\label{eq:3Drescaleddynamics}
\end{equation}
The characteristic scales normalize the base spherical or circular vesicle radius to one and absorb the viscosity into the viscous time scale, thereby removing the explicit viscosity parameter from the nondimensional Stokes equations.
The problem has three nondimensional numbers: the density ratio $\chi$, the nondimensional area-compression modulus $K_A$, and the spontaneous curvature $\meancurv_0$.
All further discussion involving the free energy and dynamics uses these nondimensional variables, and the tilde symbol is omitted henceforth for convenience.

\subsection{Relation to incompressible models}
The large-\(\Areamod\) limit recovers the inextensible membrane under a regular asymptotic scaling where $\rho\to1$ and $\gamma_{\Areamod}\to\gamma= O(1)$ as $\Areamod\to\infty$.  At finite \(\Areamod\), the parameter \(\chi\) should not be interpreted as the actual geometric excess area, because the membrane area or length is not constrained.  Rather, with the preferred density normalized to one, \(\chi\) measures the conserved lipid content:
% \begin{equation}
% \int_\Gamma \rho\,dA
% =
% \begin{cases}
% 4\pi\chi & \text{in three dimensions} \\
% 2\pi\chi, & \text{in two dimensions}.
% \end{cases}
% \end{equation}
\begin{equation}
    \int_\Gamma \rho\,dA
    =
    \begin{cases}
        4\pi\chi, & \text{in three dimensions},\\
        2\pi\chi, & \text{in two dimensions}.
    \end{cases}
\end{equation}
Thus \(4\pi\chi\) and \(2\pi\chi\) are nominal preferred area and length values: they are the values for which a spatially uniform membrane would have density \(\rho=1\).  In the regular inextensible limit (where \(\rho\to1\) and \(\gamma_{\Areamod}\to O(1)\)),  these nominal values become the actual constrained area or length, giving \(\Delta_S=4\pi(\chi-1)\) in three dimensions and \(\Delta_L=2\pi(\chi-1)\) in two dimensions.  At finite \(\Areamod\), however, \(\chi\) fixes lipid content, while the actual area or length is selected by the balance of bending, density penalty, and the imposed volume or area constraint.

In the large-\(\Areamod\) limit it is useful to distinguish the local
stretching mismatch from the geometric excess area.  Let \(\Gamma_0\) denote
the material reference surface and let $\Js=\frac{dA}{dA_0}$
be the local area stretch (where $A_0 = \int_{\Gamma_0} dA_0$ is the area of the material reference surface).  In a local material-density formulation,
conservation of membrane material gives $\rho \Js=1$ (see \refapp{appsec:FreeEnergy}).
The finite-\(\Areamod\) energy $\mathcal L_{\Areamod}[\Gamma,p]$, with the fixed volume constraint, is
\begin{equation}
    \mathcal L_{\Areamod}[\Gamma,p]
    = E - p\bigl(V[\Gamma]-V_0\bigr) \equiv
    E_{\mathrm b}[\Gamma]
    +
    \frac{\Areamod}{2}
    \int_{\Gamma}(\rho-1)^2\,dA
    -
    p\bigl(V[\Gamma]-V_0\bigr),
\end{equation}
or, equivalently,
\begin{equation}
    \mathcal L_{\Areamod}[\Gamma,p]
    =
    E_{\mathrm b}[\Gamma]
    +
    \int_{\Gamma_0}
    \frac{\Areamod}{2}\frac{(\Js-1)^2}{\Js}\,dA_0
    -
    p\bigl(V[\Gamma]-V_0\bigr).
    \label{eq:finite-KA-functional}
\end{equation}
If an independent bare surface energy is included in the model, one should add
\(\sigma_{\mathrm b}\int_\Gamma dA\) to
\eqref{eq:finite-KA-functional}.  In what follows we set
\(\sigma_{\mathrm b}=0\), so that the limiting tension is the constraint
reaction associated with inextensibility, not a residual stored surface energy.

The tension conjugate to the area stretch is
\begin{equation}
\gamma_{\Areamod}
    =
    \frac{\partial}{\partial \Js}
    \left[
        \frac{\Areamod}{2}\frac{(\Js-1)^2}{\Js}
    \right]
    =
    \frac{\Areamod}{2}\left(1-\Js^{-2}\right)
    =
    \frac{\Areamod}{2}(1-\rho^2).
    \label{eq:finite-KA-tension}
\end{equation}
For a regular inextensible limit: $\Areamod\rightarrow \infty$,  $\gamma_{\Areamod}\rightarrow \gamma=O(1)$ and $\rho\rightarrow 1$ with $\rho \Js = 1$
%requiring %\(\gamma_{\Areamod}=O(1)\).  Equivalently,
\begin{equation}
    \rho
    =
    1-\frac{\gamma}{\Areamod}
    +
    O(\Areamod^{-2}),
    \qquad
    \Js
    =
    1+\frac{\gamma}{\Areamod}
    +
    O(\Areamod^{-2}).
    \label{eq:regular-limit-rho-J}
\end{equation}
The density-penalty energy $E_{\Areamod}^{\mathrm{pen}}$ vanishes in the regular
inextensible limit
\begin{equation}
    E_{\Areamod}^{\mathrm{pen}}
    =
    \frac{\Areamod}{2}
    \int_{\Gamma}(\rho-1)^2\,dA
    =
    \frac{1}{2\Areamod}
    \int_{\Gamma_0}\gamma^2\,dA_0
    +
    O(\Areamod^{-2})
    \longrightarrow 0 .
    \label{eq:penalty-energy-vanishes}
\end{equation}
Its first variation, however, remains finite:
\begin{equation}
    \delta E_{\Areamod}^{\mathrm{pen}}
    =
    \int_{\Gamma_0}\gamma_{\Areamod}\,\delta \Js\,dA_0
    \longrightarrow
    \int_{\Gamma_0}\gamma\,\delta \Js\,dA_0 .
    \label{eq:finite-first-variation}
\end{equation}
Hence the penalty converges to the hard
constraint \(\Js=1\), while the finite limiting tension \(\gamma\) survives as
the Lagrange multiplier enforcing that constraint.
The corresponding inextensible limiting functional is therefore
\begin{equation}
    \mathcal L_{\mathrm{inc}}[\Gamma,\gamma,p]
    =
    E_{\mathrm b}[\Gamma]
    +
    \int_{\Gamma_0}\gamma(\Js-1)\,dA_0
    -
    p\bigl(V[\Gamma]-V_0\bigr).
    \label{eq:inextensible-functional}
\end{equation}
The multiplier term vanishes on admissible configurations satisfying \(\Js=1\),
but its variation gives the finite tension force in the membrane stress
balance.  Equivalently,
\begin{equation}
    \int_{\Gamma_0}\gamma(\Js-1)\,dA_0
    =
    \int_{\Gamma}\gamma\,dA
    -
    \int_{\Gamma_0}\gamma\,dA_0 .
\end{equation}
For a uniform multiplier this is \(\gamma(A-A_0)\), not simply
\(\gamma A\).  Thus the commonly written term \(\int_\Gamma\gamma\,dA\) can
be understood as the shape-dependent part of the augmented constraint
functional, or as a Legendre-transformed surface contribution; it is not the
limiting value of the quadratic density-penalty energy.

The constraint \(\Js=1\) does not mean that the limiting shape is spherical.  It
only means that the membrane is locally unstretched relative to its material
area.  %Let $R_0=\left(\frac{3V_0}{4\pi}\right)^{1/3}$ and  $\Delta_0=\frac{A_0}{R_V^2}-4\pi$, where \(A_0=\int_{\Gamma_0}dA_0\) is the material area, and the excess area \(\Delta_0\ge0\), with equality only for the sphere.  
After non-dimensionalization, $R_0=1$ and $V_0 = \frac{4\pi}{3}$.
In the regular limit of large-\(\Areamod\),
\begin{equation}
    \rho_{\Areamod}\to1,
    \qquad 
    J_{s\Areamod}\to1,
    \qquad
    V[\Gamma_{\Areamod}]\to V_0,
    \qquad
    A[\Gamma_{\Areamod}]\to 4\pi+\Delta_S .
\end{equation}
Therefore the geometric excess area of the actual shape satisfies
\begin{equation}
   \lim_{\Areamod\rightarrow \infty} \Delta[\Gamma_{\Areamod}]
    =
    \lim_{\Areamod\rightarrow \infty}\frac{A[\Gamma_{\Areamod}]}{R_V^2}-4\pi
    =
    \Delta_S .
\end{equation}
The excess area survives the inextensible limit as a geometric constraint on
the limiting shape. 
This distinction is important for spherical states.  If the membrane is forced
to remain the sphere of dimensionless volume \(V_0 = 4\pi/3\), then
\(A_{\mathrm{sph}}=4\pi\), and the uniform density is
\begin{equation}
    \rho_{\mathrm{sph}}
    =
    1+\frac{\Delta_S}{4\pi}.
\end{equation}
For fixed nonzero \(\Delta_S\),
\begin{equation}
    \gamma_{\mathrm{sph}}
    =
    \frac{\Areamod}{2}
    \left[
        1-
        \left(1+\frac{\Delta_S}{4\pi}\right)^2
    \right]
    =
    O(\Areamod),
\end{equation}
and
\begin{equation}
    E_{\Areamod,\mathrm{sph}}^{\mathrm{pen}}
    =
    \frac{\Areamod}{2}
    A_{\mathrm{sph}}
    \left(\frac{\Delta_S}{4\pi}\right)^2
    =
    O(\Areamod).
\end{equation}
Thus an exactly spherical branch with fixed \(\Delta_S\neq0\) is singular in
the large-\(\Areamod\) limit.  A regular inextensible equilibrium at fixed
excess area must instead deform away from the sphere so that
\begin{equation}
    A[\Gamma_\infty]=4\pi+\Delta_S,
    \qquad
    V[\Gamma_\infty]=V_0 .
\end{equation}

The limiting equilibrium shape is selected by the constrained bending-energy
problem
\begin{equation}
    \Gamma_\infty
    \in
    \operatorname*{arg\,min}_{\Gamma}
    \left\{
        E_{\mathrm b}[\Gamma]:
        A[\Gamma]=4\pi+\Delta_S,\;
        V[\Gamma]=V_0
    \right\}.
    \label{eq:limiting-constrained-problem}
\end{equation}
For a connected homogeneous equilibrium without external tangential forcing,
the tangential balance gives \(\nabla_s\gamma=0\), so the limiting multiplier
is constant.  The stationarity condition then takes the familiar form
\begin{equation}
    \delta E_{\mathrm b}
    +
    \gamma_0\,\delta A
    -
    p_0\,\delta V
    =
    0 .
    \label{eq:area-volume-stationarity}
\end{equation}
Thus \(\gamma_0\) is not determined by the limiting spherical shape alone.  It
is determined together with the equilibrium shape and pressure by the
area-volume constrained problem.

Finally, the earlier spherical scaling calculation should be interpreted as an
onset calculation, not as the fixed-\(\Delta_0\) inextensible limit.  If
\begin{equation}
    \chi(\Areamod)
    =
    1+\alpha\Areamod^{-p}
    +
    o(\Areamod^{-p}),
    \qquad \mbox{ for constant } p > 0, \;\;\mbox{ and }
    \alpha\neq0,
\end{equation}
then
\begin{equation}
    \gamma_0(\Areamod)
    =
    \frac{\Areamod}{2}
    \left(1-\chi(\Areamod)^2\right)
    =
    -\alpha\Areamod^{1-p}
    +
    o(\Areamod^{1-p}).
\end{equation}
Thus \(p>1\) gives \(\gamma_0\to0\), \(p=1\) gives
\(\gamma_0\to-\alpha\), and \(p<1\) gives a singular limiting tension that corresponds to forcing the density mismatch to vanish
too slowly relative to the stiffness as in the case of a sheared inextensible membrane (cf. Eq.~(9.58) in \cite{vlahovska2016dynamics}).  This
path has vanishing spherical excess area, however, and therefore does not
represent a regular limit at fixed nonzero \(\Delta_S\).

\section{Spectrum and stability of spherical/circular shape}\label{sec:spectrum}
The present formulation is a finite-compressibility extension of the classical inextensible vesicle theory, rather than a replacement on physical grounds. In the classical small-deformation theory \cite{seifert1999fluid,olla2000behavior,Misbah:2006,vlahovska2007dynamics}, the membrane tension is a Lagrange multiplier determined by enforcing local area incompressibility and the fixed excess area constraint, while the hydrodynamic and kinematic conditions are expanded perturbatively about the spherical base state. This is a consistent asymptotic description when the area modulus is large and true areal dilation is negligible. In the present model, the constraint is relaxed: lipid density is evolved explicitly, and the membrane tension is obtained constitutively from density variations. This formulation is therefore more appropriate when finite area compressibility, lipid-density redistribution, or density-driven tension is part of the mechanism of interest.

\subsection{Spherical membrane\label{subsec:spherical_membrane}}
\subsubsection{Parametrization of a nearly spherical interface}
The interface of a nearly spherical membrane $\bx_\surf$ and the lipid areal density distribution   $\rho$ are parametrized in spherical coordinates as
\begin{equation}
    \bx_\surf(\theta,\phi) = \big(R+\eps f(\theta,\phi)\big)\rhat, \quad \rho(\theta,\phi) = \rho_0+ \eps g(\theta,\phi)  \label{eq:3Dparametrization}
\end{equation}
respectively where the average radius $R$ and average density $\rho_0$ are chosen to preserve the volume of the enclosure  and total lipid count respectively under deformation:
\begin{equation}
    R = 1 -\frac{1}{4\pi}\sum_{n=2}^\infty \eps^n \Delta_{V,n},\quad \rho_0 = \chi-\frac{1}{4\pi}\sum_{n=2}^\infty\eps^n\Delta_{\rho,n}\label{eq:3Dparaverage}
\end{equation}
with the corrections $\Delta_{V,n}$ and $\Delta_{\rho,n}$ given in \refeqn{eqapp:3Davgcorrections}.
%The first two non-zero corrections of order $\eps^2$ and $\eps^3$ to the average radius are given by
% \begin{equation}
%     \Delta_{V,2} = \int_{\surf_0} f^2 \,d\surf_0,\quad\Delta_{V,3} = \int_{\surf_0} f^3\,d\surf_0\label{eq:3Dvolcorrections}
% \end{equation}
% where $\Gamma_0$ is the integral over the unit sphere.
% The phospholipid density may vary across the interface and is parametrized as
% \begin{equation}
%     \rho = \rho_0+ \eps g(\theta,\phi)\label{eq:3Ddensityparametrization}
% \end{equation}
% where the baseline density $\rho_0$ is chosen to preserve the total concentration, $4\pi\rhobar$, of the membrane:
% \begin{equation}
%     \rho_0 = \rhobar+\frac{1}{4\pi}\sum_{n=2}^\infty \eps^n\Delta_{\rho,n}\label{eq:3Ddensitycorrection}
% \end{equation}
% There are corrections to the density from both the density and shape perturbations.
% The first non-zero corrections at order $\eps^2$ and $\eps^3$ and are given by
% \begin{equation}
%     \Delta_{\rho,2} = \int_{\surf_0} -f^2+\frac{1}{2}(\bnab_0f)\cdot(\bnab_0 f)+2fg\, \,d\surf_0,\quad \Delta_{\rho,3} = \int_{\surf_0} 2f^3-\frac{1}{2}g(2f+(\bnab_0f)\cdot(\bnab_0f)\,d\surf_0
% \end{equation}
% where $\bnab_0$ is the surface gradient over a unit sphere. 
The perturbation in the shape and density is further decomposed in terms of scalar spherical harmonic functions (see \refapp{app:sphericalharmonics} for definitions)
\begin{equation}
    f(\theta,\phi) = \sumjm \fjm\Yjm(\theta,\phi) ,\quad g(\theta,\phi) = \sumjm \gjm\Yjm(\theta,\phi),\label{eq:3Dsphform}
\end{equation}
where $\sumjm = \sum_{j=1}^\infty\sum_{m=-j}^j$.
The volume and total concentration corrections in \refeqn{eq:3Dparaverage}  lead to the omission of $f_{0,0}$ and $g_{0,0}$ terms in \refeqn{eq:3Dsphform}.
The membrane is centered at the origin and thus $f_{1,m}=0$ for $|m|\leq 1$. 
There is no such restriction for $g_{1,m}$ for $|m|\leq 1$.
The linearization of the tension for the boundary condition requires that $\chi-1\ll 1$.
%The representation of the hydrodynamic and interfacial stress as well as other relevant values in terms of the parametrization given in \refeqn{eq:3Dsphform} can be found in \refapp{subapp:interfacialstresses}.

\subsubsection{Solution to Stokes equation}
% Given parametrization \refeqn{eq:3Dparametrization} for the surface and density, the non-isotropic component of the Helfrich bending stress and stress from density driven tension at leading order in $\eps$
% \begin{equation}
%     \begin{split}
%         &\tau^B = \eps\frac{\Bendmod}{2}(j-1)(j+2)\big(2H_0(H_0+2)+j(j+1)\big)\fjm\Yjmtwo\\
%         &\tau^E = \Areamode\eps\bigg[\big(-2(\rhobar-1)(j-1)(j+2)\fjm-2\gjm\big)\Yjmtwo +\sqrt{j(j+1)}\gjm\Yjmzero\bigg]
% \end{split}\label{eq:3Dstressessph}
% \end{equation}
% where it has been assumed that $(\rhobar-1)\sim O(1)$.
% The isotropic components of the stress, leading to constant pressure solutions inside and outside the membrane, do not drive flow or shape deformation and have been removed.
% The $\Yjmtwo$ component of the stresses are normal to the surface of a unit sphere and govern the shape dynamics for nearly spherical shapes.
% The $\Yjmzero$ components are tangential to the unit sphere and drive redistribution of the lipid density on the interface.

The flow outside the membrane is given as a sum of decaying basis elements of the Stokes equation and the flow inside is given by a sum of growing basis elements of the Stokes equation
\begin{equation}
    \bu^\out = \sumjmq c_{j,m,\q}\bu_{j,m,\q}^-, \quad\bu^\ins = \sumjmq c_{j,m,q}\bu_{j,m,\q}^+. \label{eq:3Dsolansatz}
\end{equation}
where $\bu_{j,m,\q}^\pm$ comprise a basis for the Stokes equation specified in \refeqn{eqapp:3Dstokesbasis}. 
The form given in \refeqn{eq:3Dsolansatz}, satisfies the conditions for continuity of flow at the interface, \refeqn{eq:3DBC1}, and decay of the flow as $|\bx|\to \infty$ , \refeqn{eq:3DBC2}.
The remaining stress balance conditions, \refeqn{eq:3DBC3}, will specify the value of coefficients $c_{j,m,\q}$ by solving for them in the stress balance equations 
\begin{equation}
    \tau^{\HD}_{j,m,\q}=\tau^B_{j,m,\q}+\tau^E_{j,m,\q}.\label{eq:stressbalancesph}
\end{equation}
where the hydrodynamic stress, $\tau^{\HD}$ is given in \refeqn{eqapp:HDstress} and the interfacial stresses $\tau^B$ and $\tau^E$ are given in \refeqn{eqapp:3Dstressessph}.
For the three-dimensional calculation of the spectrum, the tension has been regularized with respect to preferred density near 1: $\chi-1\ll 1$.
The resulting coefficients $c_{j,m,\q}$ are given in \refeqn{eqapp:3Dcoefsol} and \refeqn{eqapp:3Dcoefsol2}.
% The leading order hydrodynamic stress is given by 
% \begin{equation}
%     \tau^{\HD} = \sumjmq \tau_{j,m,\q}^{\HD,\pm} \by_{j,m,\q}.
% \end{equation}
% where $\tau_{jm\q}^{\HD,\pm} = \sum_{\q'=0}^2 c_{jm\q'}^\pm T^\pm_{\q\q'}$ and $T^\pm_{\q\q'}$ are given in \refapp{app:sphericalharmonics}. 
% Solving the resulting equation $\tau^{\HD} = \tau^B+\tau^E$ gives
% \begin{equation}
%     \begin{split}
%     &c_{j,m,0} = \eps\big(\sqrt{ j(j+1)}\xi_{j,m}\big)^{-1}\bigg[ p_{j,m,0}^f\fjm+p_{j,m,0}^g\gjm\bigg] \\
%     &c_{j,m,1} =0 ,\quad
%     c_{j,m,2} = \eps\xi^{-1}_{j,m}\bigg[ p_{j,m,2}^f\fjm+p_{j,m,2}^g\gjm\bigg] 
%     \end{split}\label{eq:3Dcoefsol}
% \end{equation}
% where
% \begin{equation}
% \begin{split}
%     &\xi = \visc(2j+1)(2j-1)(2j+3)\\
%     &p_{j,m,0}^f = \frac{3}{2}(j-1)j(j+1)(j+2)(2j+1)\big(2\Areamod(\rhobar-1)-\Bendmod(2\meancurv_0(\meancurv_0+2)+j(j+1))\big)\\
%     &p_{j,m,2}^f = j(j-1)(j+1)(j+2)(2j+1)(2\Areamod(\rhobar-1)-\Bendmod\big(2\meancurv_0(\meancurv_0+2)+j(j+1)\big)\big)\\
%     &p_{j,m,0}^g=-\Areamod j(j+1)(2j^2+2j-3),\quad p_{j,m,2}^g = \Areamod j(j+1)
%     \end{split}.
% \end{equation}
The shape evolution, \refeqn{eq:3Dshapeevo}, and density evolution, \refeqn{eq:3Ddensevo}, for each mode at leading order in $\eps$ are given by
\begin{equation}
    \eps\frac{d\fjm}{dt} = c_{j,m,2}, \quad \eps\frac{d\gjm}{dt} = -2c_{j,m,2}+\sqrt{j(j+1)}c_{j,m,0}. \label{eq:3Dshapedensityevosph}
\end{equation}
\refeqn{eq:3Dshapedensityevosph} together with \refeqn{eqapp:3Dcoefsol}  and \refeqn{eqapp:3Dcoefsol2} gives for $j\geq 2$
\begin{equation}
    \begin{split}
    &\frac{d}{dt}\bmat{\fjm\\\gjm} = \bP \bmat{\fjm\\\gjm},\quad \bP =\xi(j)^{-1} \bmat{B_1(j)\zeta(j)&B_2(j)\Areamod\\-\frac{1}{2}B_1(j)\zeta(j)&-B_2(j)\Areamod (2j^2+2j-1)}
    \end{split}\label{eq:3Dsdevo}
\end{equation}
where 
\begin{equation}
    \begin{split}
    &B_1(j)=(j-1)j(j+1)(j+2),\quad B_2(j)=j(j+1)\\
    &\zeta(j) = -\big(j(j+1)+2\meancurv_0(\meancurv_0+2)\big)+2\Areamod (\chi-1), \quad\xi(j) = (2j+1)(2j-1)(2j+3)
    \end{split}\label{eq:3Dzetadef}
\end{equation}
Note $B_1(j)$, $B_2(j)$ and $\xi(j)$ are positive for all $j\geq 2$.
The determinant and trace are
\begin{equation}
    \begin{split}
        &\det(\bP) = -\frac{\Areamod}{2}\xi(j)^{-2}B_1(j)B_2(j)(4j^2+4j-3) \zeta(j)\\
        &\text{tr}(\bP) = \xi(j)^{-1}(B_1(j)\zeta(j)-B_2(j)(2j^2+2j-1))
    \end{split}.
\end{equation}
The stability of the $(j,m)$ shape and density modes are determined by $\zeta(j)$. When $\zeta(j)<0$, the determinant is positive and the trace is negative indicating two negative eigenvalues and a stable spherical fixed shape. When $\zeta(j)>0$, the determinant is negative and the sphere is a saddle point. 
The eigenvalues are 
\begin{equation}
    \nu_j^\pm=\frac{1}{2}\bigg(\text{tr}(\bP)\pm\sqrt{\text{tr}(\bP)^2-4\det(\bP)}\bigg)
    \label{eq:eigenvalues3D}
\end{equation}
For each shape mode $j\geq 2$, expanding about $\zeta(j)=0$, the eigenvalue and vector pairs for the evolution of the $j$-th shape and density modes, \refeqn{eq:3Dsdevo}, up to order $O(\zeta^2)$ are
\begin{equation}
\begin{split}
    &\bigg\{- \frac{\Areamod (2j^2+2j-1)B_2(j)}{\xi(j)}+\frac{B_1(j)\zeta(j)}{2\xi(j)(2j^2+2j-1)},\bmat{-\frac{1}{2j^2+2j-1}+\zeta(j)\frac{(4j^2+4j-3)B_1(j)}{2\Areamod B_2(j)(3j^2+2j-1)^3}\\1}\bigg\}\\
    &\bigg\{\zeta(j)\frac{(4j^2+4j-3)B_1(j)}{2(2j^2+2j-1)\xi(j)},\bmat{-\frac{2\Areamod B_2(j)(2j^2+2j-1)}{B_1(j)}-\zeta(j)\frac{4j^2+4j-3}{2j^2+2j-1}\\\zeta(j)}\bigg\}
    \end{split}
\end{equation}
The first eigenvalue is negative near the bifurcation at $\zeta(j) = 0$. 
The second eigenvalue changes signs as $\zeta(j)=0$ is crossed and the eigenvector direction is strictly in the shape only as $\zeta(j)\to0$.
Due to the relation $\zeta(j)>\zeta(j')$ for $j>j'$, the stability of the sphere will be determined by the stability of the $j=2$ shape mode and the criteria for the stability of the sphere will be $\zeta(2)<0$.

{\color{black}
The negative eigenvalue also gives a useful interpretation of density relaxation. Since this mode is primarily a density perturbation near $\zeta(j)=0$, its leading-order decay rate is
\begin{equation}
\nu_{\rm fast}
=
-\frac{\Areamod j(j+1)\left(2j(j+1)-1\right)}{\xi(j)}
+O(\zeta).
\end{equation}
Equivalently, writing $\Lambda_j=j(j+1)$,
\begin{equation}
\nu_{\rm fast}
=
-\Lambda_j\frac{\Areamod(2\Lambda_j-1)}{\xi(j)}
+O(\zeta).
\end{equation}
This decay has the same modal structure as surface diffusion on a sphere, for which the decay rate of a spherical-harmonic density mode is proportional to $-\Lambda_j D$. Thus the coupled membrane--Stokes dynamics defines the nondimensional mechanics-induced density-relaxation scale
\begin{equation}
D_{\rm mech}^{\rm nd}(j)
=
\frac{\Areamod(2\Lambda_j-1)}{\xi(j)}.
\label{eq:mechD}
\end{equation}
For the first nontrivial spherical mode $j=2$, $\Lambda_j=6$ and $\xi(2)=105$, so
\begin{equation}
D_{\rm mech}^{\rm nd}(2)=\frac{11}{105}\Areamod.
\end{equation}
In dimensional variables,
\begin{equation}
D_{\rm mech}(2)=\frac{11}{105}\frac{K_A^{\rm phys}R_0}{\mu},
\end{equation}
where $K_A^{\rm phys}$ is the dimensional area-compression modulus. Taking $K_A^{\rm phys}\sim0.2\,{\rm N/m}$, $R_0\sim1\text{--}10\,\mu{\rm m}$, and $\mu\sim10^{-3}\text{--}5\times10^{-2}\,{\rm Pa\,s}$ gives
\begin{equation}
D_{\rm mech}(2)\sim10^5\text{--}10^8\,\mu{\rm m}^2/{\rm s},
\end{equation}
which is much larger than the molecular lateral diffusivity of fluid lipids, $D_{\rm mol}\sim1\text{--}10\,\mu{\rm m}^2/{\rm s}$, with measurements in free-standing DOPC giant unilamellar vesicles giving
\(D_{\rm mol}\simeq 7\text{--}8,\mu{\rm m}^2/{\rm s}\), and classical planar-bilayer
measurements giving comparable values above the lipid melting transition
\cite{Fahey1977LateralDiffusion,Doeven2005GUVLateralMobility,Przybylo2006LipidDiffusionGUV}.
% Molecular lateral diffusion coefficients of lipids in fluid vesicle membranes are typically of order
% \(D_{\rm mol}\sim 1\text{--}10,\mu{\rm m}^2/{\rm s}\), with fluorescence-correlation spectroscopy measurements in free-standing DOPC giant unilamellar vesicles reporting values near
% \(7\text{--}8,\mu{\rm m}^2/{\rm s}\)
% \cite{Doeven2005GUVLateralMobility,Przybylo2006LipidDiffusionGUV}.
The corresponding mechanics-based Peclet number,
\begin{equation}
{\rm Pe}_{\rm mech}=\frac{U_sL_s}{D_{\rm mech}},
\end{equation}
is therefore typically much smaller than unity for vesicle-scale membrane motions. 
Thus, for the low-order vesicle-scale modes considered here, density redistribution is dominated by the mechanics-induced relaxation already contained in the coupled membrane–Stokes equations; molecular diffusion would primarily provide an additional short-wavelength regularization.
%Therefore, in the parameter regime considered here, density redistribution is dominated by the mechanics-induced relaxation already contained in the coupled membrane--Stokes equations.
}

The coordinate system is chosen about the center of mass of the fluid enclosed by the membrane. 
As a result $f_{1,m}=0$, the velocity of the center of mass of the droplet at linear order is instead given as \cite{kawakami2025migration}
%As a result instead of an evolution equation for $f_{1,m}$, which must remain 0 for the coordinate system to remain at the center of mass, the velocity of the center of mass is computed.
%The motion of the center of mass \cite{kawakami2025migration} at linear order depends only on the $j=1$ mode for the density distribution $g_{1,m}$:
\begin{equation}
    \bU_\drop = \frac{\Areamod}{5\sqrt{6\pi}}\bmat{(g_{1,-1}-g_{1,1})\\-i(g_{1,-1}+g_{1,1}) \\\sqrt{2}g_{1,0}}.
\end{equation}
The evolution of the density for the $j=1$ mode follows the same form as \refeqn{eq:3Dshapedensityevosph} with the reduced version given by
\begin{equation}
    \frac{dg_{1,m}}{dt} = -\frac{2}{5}\Areamod g_{1,m}.\label{eq:gj1evoeq}
\end{equation}
Consequently, in the stable state, \(g_{1m}=0\) for all \(|m|\le1\), and the drop velocity \(U_d=0\).
% As a result for the stable drop velocity is $\bU_\drop = 0$, with corresponding density $g_{1,m} = 0$ for all $|m|\leq 1$. 

\subsection{Circular membrane}

The circular problem has a useful simplification: in two dimensions, lipid conservation along material arcs allows the density to be written directly in terms of the local arclength stretch. This reduces the density dynamics to a shape-dependent contribution to the energy. We summarize the main identities and spectral consequences here; the detailed shape variations and polar-coordinate Stokes calculation are collected in \refapp{app:2Dmodels}.

\subsubsection{Energy identity and interfacial condition}
Let $\Gamma(t)$ be a closed curve parameterized by the Lagrangian coordinate $\theta\in\mathbb S$ through $\bm X(\theta,t)$, and set
\begin{equation}
    Q(\theta,t)=\left|\partial_\theta \bm X(\theta,t)\right| .
\end{equation}
Lipid conservation on each material arc gives
\begin{equation}
    \frac{d}{dt}\left(\rho(\bm X(\theta,t),t) Q(\theta,t)\right)=0.
\end{equation}
For a uniformly distributed material lipid mass, the nondimensional density is therefore slaved to the shape,
\begin{equation}
    \rho(\bm X(\theta,t),t)=\frac{\chi}{Q(\theta,t)},
\end{equation}
where $\chi>1$ is the nondimensional total lipid mass per unit reference arclength. With the bending rigidity rescaled to one, the total nondimensional energy becomes
\begin{align}
    E[\Gamma]
    &=\frac{1}{2}\int_\Gamma (\meancurv-\meancurv_0)^2+\Areamod(\rho-1)^2\,ds\nonumber\\
    &=\frac{1}{2}\int_{\mathbb S}\frac{P(\bm X)^2}{Q(\bm X)^5}
      +\meancurv_0^2 Q(\bm X)
      +K_A\frac{(Q(\bm X)-\chi)^2}{Q(\bm X)}\,d\theta
      -2\pi \meancurv_0,
\end{align}
where
\begin{equation}
    P(\bm X)=R\partial_\theta\bm X\cdot\partial^2_\theta\bm X,
    \qquad
    R=\begin{bmatrix}0&1\\-1&0\end{bmatrix}.
\end{equation}
We assume that the enclosed area is fixed as $\pi$. The interfacial force in the Stokes traction balance is
\begin{equation}
    \jump{\bm\sigma(\bm u,p)\bm n}
    =-\frac{1}{Q}\left(
    \frac{\delta E_1}{\delta\bm X}
    +\meancurv_0^2\frac{\delta E_2}{\delta\bm X}
    +K_A\frac{\delta E_3}{\delta\bm X}
    \right),
\end{equation}
where $E=E_1+\meancurv_0^2E_2+K_AE_3-2\pi \meancurv_0$. The explicit variational derivatives are given in \refapp{app:2Dmodels}.

\subsubsection{Linearized modal spectrum}
The unit circle is a steady solution with pressure jump
\begin{equation}
    \jump{p_0}=\frac{-1+\meancurv_0^2+K_A(1-\chi^2)}{2}.
\end{equation}
Linearizing about this state and eliminating the Stokes flow gives a closed system for each Fourier mode,
\begin{equation}
    \frac{d}{dt}\begin{bmatrix}Y_m^r\\Y_m^\theta\end{bmatrix}
    =\bm M_m\begin{bmatrix}Y_m^r\\Y_m^\theta\end{bmatrix}.
\end{equation}
For $m=0$, $\bm M_0$ is the zero matrix. For $m=1$,
\begin{equation}
    \bm M_1=
    \begin{bmatrix}
    -\frac{K_A\chi^2}{8} & -i\frac{K_A\chi^2}{8}\\
    i\frac{K_A\chi^2}{8} & -\frac{K_A\chi^2}{8}
    \end{bmatrix},
\end{equation}
with eigenvalues $0$ and $-K_A\chi^2/4$. The zero eigenvalue corresponds to rigid translation of the circle, while the second eigenvalue is stable.
For $m\ge 2$,
\begin{equation}
\bm M_m=
\begin{bmatrix}
-\frac{m\left(2m^2+K_A(1-\chi^2)+(\meancurv_0^2-3)\right)}{8} & 0\\
-i\frac{2m^2+K_A(1-3\chi^2)+(\meancurv_0^2-3)}{8} & -\frac{mK_A\chi^2}{4}
\end{bmatrix}.
\end{equation}
The corresponding eigenvalues are
\begin{equation}
    \nu_1=-\frac{K_A\chi^2}{4}m,
    \qquad
    \nu_2=-\frac{m}{8}\left(2m^2+K_A(1-\chi^2)+(\meancurv_0^2-3)\right).
\end{equation}
The first eigenvalue is always stable for $K_A>0$ and $\chi>0$. The second eigenvalue changes sign when the density-driven reduction of the energy overcomes bending stabilization. Thus the circular membrane is unstable in mode $m\ge2$ when
\begin{equation}
    K_A(\chi^2-1)>(2m^2-3)+\meancurv_0^2 .
\end{equation}
The first nontrivial instability is therefore the $m=2$ shape mode.

\section{Shape bifurcation at onset of instability of circular/spherical shape}\label{sec:variational}
We compute variations of the energy functional with respect to shape and density to examine the onset bifurcations of circular membranes in two dimensions and spherical membranes in three dimensions.
% Variations of the energy functional due to shape and density variation is computed to examine the shape bifurcation of the circular shape in two dimensions and the spherical shape in three dimensions.

\subsection{Stability of a spherical membrane in three-dimensional space}

Under parametrization in \refeqn{eq:3Dparametrization}, the free energy of a perturbed sphere is 
\begin{equation}
    E = \sum_{n=0}^\infty \eps^n E^{(n)}.
\end{equation}
$E^{(0)}$ corresponds to the free energy of the undeformed sphere.
The first variation of the energy vanishes, %\col{(Po: In (4.2), is it $K_A$?)}
\begin{equation}
     E^{(1)} = \sumjm\int_\surf (\meancurv_0+1)(2\meancurv_0+j(j+1))\fjm+\Areamod(\chi-1)(\gjm+(\chi-1)\fjm) d\surf = 0
\end{equation}
confirming that the sphere is a fixed shape for all choices of $\Areamod,\chi,\meancurv_0$. 
The second variation is 
\begin{equation}
\begin{split}
     &E^{(2)} = \sumjm\bmat{\fjm^\ast,\gjm^\ast}\bM_{j,m}\bmat{\fjm\\\gjm}\\
     &\bM_{j,m} = \bmat{\frac{(j-1)(j+2)}{8}[(2\meancurv_0(\meancurv_0+2)+j(j+1))-2\Areamod(\chi^2-1)]&0\\0&\frac{1}{2}\Areamod}
     \end{split}
\end{equation}
The eigenvalues of $\bM_{j,m}$ are $-\lambda^f(j)$ and $-\lambda^g(j)$, where $\lambda^f(j)=-\frac{(j-1)(j+2)}{8}[(2\meancurv_0(\meancurv_0+2)+j(j+1))-2\Areamod(\chi^2-1)],\lambda^g(j)=-\frac{1}{2}\Areamod$. 
$\lambda^g(j)<0$ indicates an increase of free energy when density perturbations are present and as a result isotropic density of phospholipids is favored.
When $\lambda^f(j)$ is negative the spherical shape is energetically favored.
Instability of the spherical shape occurs when $\lambda^f(j)>0$.
It can be observed that there are competing effects of the Helfrich bending energy, stabilizing the spherical shape, and the density component, driving the shape to have larger surface area than the base sphere.
Under the assumption $\chi-1\ll1$, $\lambda^f(j) =\frac{(j-1)(j+2)}{8}\zeta(j)+O(\chi-1)$ where $\zeta(j)$ \refeqn{eq:3Dzetadef} is the bifurcation parameter obtained via dynamical equations in \refsec{sec:spectrum}.
The difference is due to the linearization of the tension in the Stokes equation formulation (\refeqn{eq:elasticstress2}).
The critical area modulus, $(\Areamod)_{j}^\ast$, for the destabilization of each mode for fixed $\meancurv_0$ and $\chi$ is
\begin{equation}
    (\Areamod)_{j}^\ast = \frac{2\meancurv_0(\meancurv_0+2)+j(j+1)}{2(\chi^2-1)}.\label{eq:3Dcritareamod}
\end{equation}
In the limit of $\Areamod \to 0$, the spherical shape is stable for all modes. 
As $\Areamod$ increases, the $j=2$ shape mode will destabilize first at $\Areamod = (\meancurv_0^2+2\meancurv_0+3)/(\chi^2-1)$.

\subsubsection{Emergence of non-spherical shapes and bifurcation at critical area modulus}
We explore the bifurcation of the steady shape around the instability of the second shape mode occurring at $\Areamod=(\Areamod)_2^\ast$. 
From the linear analysis, it is known that all other shape modes with index $j>2$ are stable in the unperturbed state, so it is sufficient to consider only shapes with the $j=2$ mode perturbed.
Furthermore we consider only axisymmetric shapes: 
\begin{equation}
    \bx_\surf = (R+\eps f_{2,0}\Yspp_{2,0}(\theta,\phi))\rhat\label{eq:3Dj2shape}
\end{equation}
and the density is assumed to be uniform.

The free energy of the membrane with shape \refeqn{eq:3Dj2shape} is 
\begin{equation}
    \begin{split}
        &E = E_0+\eps^2 E_2+\eps^3 E_3+\eps^4 E_4+O(\eps^5)\\
        &E_2 = -\lambda^f(2)f_{2,0}^2,\quad E_3 = \sqrt{\frac{5}{\pi}}\frac{2\Areamod(5\meancurv_0-3)(\chi^2-1)+(\meancurv_0^2-8\meancurv_0+9)\lambda^f(2)}{21(\meancurv_0^2+2\meancurv_0+3)}f_{2,0}^3\\
        &E_4 = \frac{\Areamod\left(8\meancurv_0-27+\chi^2(7\meancurv_0^2+6\meancurv_0+48)\right)+\lambda^f(2)(19\meancurv_0^2+54\meancurv_0+3)}{28(\meancurv_0^2+2\meancurv_0+3)\pi}f_{2,0}^4
    \end{split}
\end{equation}
At the bifurcation point $\lambda^f(2) = 0$, $E_3$ is non-zero when $\meancurv_0\neq \frac{3}{5}$ and there is a transcritical bifurcation.
The sign of the third variation matches that of $5\meancurv_0-3$ indicating a preference at the bifurcation point for prolate shapes when $\meancurv_0<\frac{3}{5}$ and oblate shapes when $\meancurv_0>\frac{3}{5}$. 
The threshold of $\meancurv_0=\frac{3}{5}$ has previously been observed incompressible membranes \cite{zhong1989bending}. Note there is a factor of $-1/2$ difference in the definition of mean curvature.

\begin{figure}
    \centering
    \includegraphics[width=\linewidth]{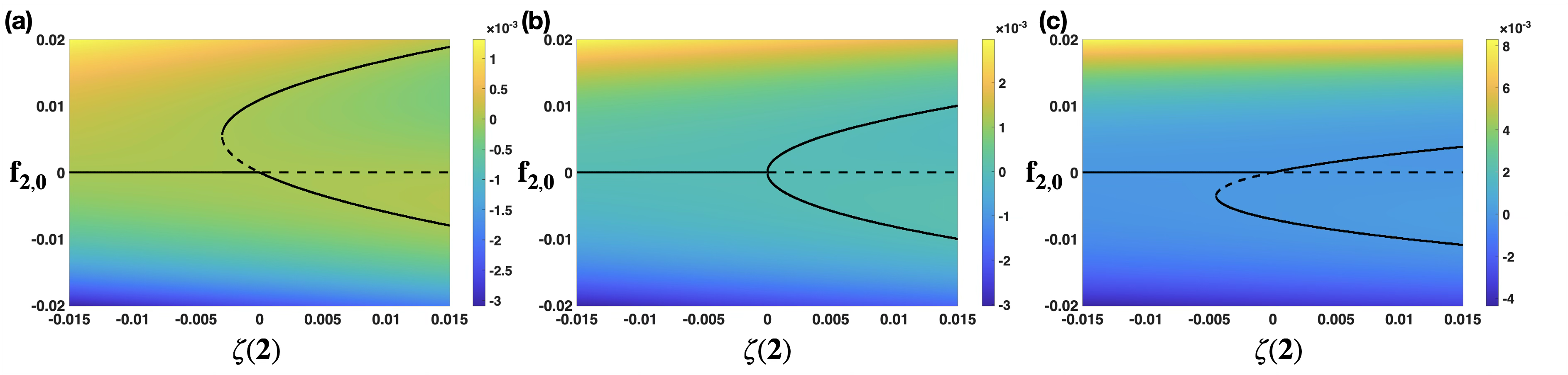}
    \caption{Bifurcation diagram for three-dimensional membrane near onset of spherical shape instability. Axes correspond to bifurcation parameter $\zeta(2)$ (\refeqn{eq:3Dzetadef}) and shape $f_{2,0}$ near bifurcation point. Color indicates $\frac{\partial E}{\partial f_{2,0}}$. Solid line indicates stable shapes and dotted lines indicate unstable. For both plots $\chi=1.01$. (a): $\meancurv_0=0<3/5$, leading to transcritical bifurcation with prolate shape preference (b): $\meancurv_0=3/5$, leading to pitchfork bifurcation (c): $\meancurv_0=3>3/5$, leading to transcritical bifurcation with oblate shape preference.}
    \label{fig:bifurcationplot}
\end{figure}

Near the bifurcation point when $\lambda^f(2) \sim\eps$, balance between the second and third variation, $E_2+\eps E_3 = 0$, gives shape solution 
\begin{align}
    f_{2,0} &=\sqrt{\frac{\pi}{5}}\frac{21(\meancurv_0^2+2\meancurv_0+3)\lambda^f(2)}{(\meancurv_0^2-8\meancurv_0+9)\lambda^f(2)+2(5\meancurv_0-3)(\chi^2-1)} \\
    &= \sqrt{\frac{\pi}{5}}\frac{21(\meancurv_0^2+2\meancurv_0+3)}{2(5\meancurv_0-3)(\chi^2-1)} \lambda^f(2)+O\left(\lambda^f(2)^2\right).\label{eq:3Dtranscirticcalslope}
\end{align}
An illustration of the energy landscape and the bifurcation for the parameter $\meancurv_0=0,\chi=1.01$ with variable $K_A$ is given in \reffig{fig:bifurcationplot}.
%, with \ref{eq:3Dtranscirticcalslope} given as a dotted line.
The x-axis is given by the $\zeta(2) \sim -\lambda^f(2)$, introduced in \refeqn{eq:3Dzetadef}, which is a monotonically increasing function of the area modulus $\Areamod$ and preferred density $\chi$, with the shape bifurcation occurring at $\zeta(2) = 0$.
For low area modulus $K_A$ (small $\zeta(2)<0$), the spherical shape is stable and is the only fixed shape. 
In \reffig{fig:bifurcationplot} the bifurcation point is approached, the system becomes bi-stable, with a secondary prolate shape becoming stable for $\meancurv_0<\frac{3}{5}$ and the oblate shape for $\meancurv_0>\frac{3}{5}$ before reaching the bifurcation point.
As the bifurcation point is crossed, the system remains bi-stable with a stable prolate and oblate fixed shapes while the spherical shape becomes unstable.
As predicted by the energy variation, the bifurcation is a transcritical type and the slope of the bifurcation near $\zeta(2)=0$ is determined by the spontaneous curvature $\meancurv_0$.
For $\meancurv_0=3/5$, where the bifurcation transitions from prolate to oblate favoring, the bifurcation type is pitchfork.

Many simulations of inextensible membranes have used a global area constraint to enforce a fixed excess area condition. 
The emergence of the stable prolate shape as we cross the bifurcation point is observed in computational models such as those in \cite{hollo2021shape}. Our model also predicts that if the spontaneous curvature is set to $\meancurv_0>\frac{3}{5}$, a stable oblate shape will instead emerge.
%\col{Explain the $\rho^\ast$ connection and why the prolate shape appears}

When $\meancurv_0=\frac{3}{5}$, the third variation vanishes. 
The fourth variation is given by
\begin{equation}
    E_4|_{\meancurv_0=3/5} = \Areamod\frac{451\chi^2-185}{532\pi}f_{2,0}^4
\end{equation}
$E_4>0$ in the parameter regime $\chi>1$, thus there is a supercritical pitchfork bifurcation.
In three dimensions, the asymmetry of the prolate and oblate shapes, which is not present in two dimensions, results in a transcritical bifurcation instead of a pitchfork bifurcation.

\subsection{Stability of circular membrane in two-dimensional space}
We now consider the two-dimensional case, for which the steady state is the unit circle. Parameterizing the interface by Lagrangian coordinates $\bm X(\theta,t)$, the density advection law reduces to an ODE along material points,
\begin{equation}
    \frac{d}{dt}\paren{\rho(\theta,t)\abs{\partial_\theta\bm X(\theta,t)}} = 0.
\end{equation}
As a consequence, the density is slaved to the local stretching of the parameterization via
$\rho=\chi/\abs{\partial_\theta\bm X(\theta,t)}$, where $\chi$ is the conserved nondimensional lipid mass per unit reference arclength. 
With the bending rigidity rescaled to one, the two-dimensional energy is
\begin{equation}
\begin{aligned}
    E[\bm X]
    &= \frac{1}{2}\int_\Gamma
    \paren{H-\meancurv_0}^2+K_A\paren{\rho-1}^2\,ds\\
    &= \frac{1}{2}\int_{\mathbb S}
    \frac{\abs{R\partial_\theta\bm X\cdot\partial_\theta^2\bm X}^2}
    {\abs{\partial_\theta\bm X}^5}
    +\meancurv_0^2\abs{\partial_\theta\bm X}
    +K_A
    \frac{\paren{\abs{\partial_\theta\bm X}-\chi}^2}
    {\abs{\partial_\theta\bm X}}\,d\theta
    -2\pi \meancurv_0.
\end{aligned}\label{eq:2Dvariational_energy}
\end{equation}
The rotation matrix $R$ is
\begin{equation}
    R = 
    \begin{bmatrix}
        0 & 1 \\
        -1 & 0
    \end{bmatrix}.
\end{equation}
Fixing $m\in\mathbb Z$ with $|m|\ge 1$, we perturb the unit circle in the eigenvector direction,
\begin{equation}
\bm X(\theta)
=(1+\delta_\varepsilon)\,\ebr(\theta)
+\varepsilon\Big(\cos(m\theta)\,\ebr(\theta)-\frac1m\sin(m\theta)\,\ebtheta(\theta)\Big),
\qquad \theta\in[0,2\pi].
\end{equation}
Here $\delta_\varepsilon=\frac{m^2-1}{4m^2}\varepsilon^2-\frac{(m^2-1)^2}{32m^4}\varepsilon^4+\mathcal O(\varepsilon^6)$ is chosen to enforce the fixed-area constraint $A(\varepsilon)=\pi$. The perturbation is chosen so that, to leading order, it corresponds to an $m$-fold shape deformation rather than a mere reparameterization. Because the perturbation is rotation invariant, all odd powers of $\varepsilon$ integrate to zero; hence,
\begin{equation}
E=E_0+E_2\varepsilon^2+E_4\varepsilon^4+\mathcal O(\varepsilon^6).
\end{equation}
The zeroth order term gives the energy of the steady state,
\begin{equation}
E_0=\pi\Big(1+\meancurv_0^2+K_A(\chi-1)^2\Big)-2\pi \meancurv_0.
\end{equation}
The second variation is
\begin{equation}E_2=-\frac{\pi}{4}\,(m^2-1)\zeta(m), \quad \zeta(m) = -
(\meancurv_0^2+2m^2-3) +K_A(\chi^2-1).
\end{equation}

For $m=\pm 1$, the second variation vanishes for all parameter regimes, so there is no energetic restoring force in these directions. These modes correspond to rigid motions of the circle (translations/rotations), which leave both curvature and local stretching unchanged to this order.
For $\chi>1$ and $|m|\ge 2$, the factor $K_A(1-\chi^2)$ is destabilizing: increasing surface length at fixed enclosed area dilutes the lipids (lowering the density penalty) and can outweigh the bending cost associated with higher curvature. Consequently, the $m$-th mode becomes unstable when $K_A>K_A^*$, where the critical value is
\begin{equation}
    K_A^* = \frac{\meancurv_0^2+2m^2-3}{\chi^2-1}.
\end{equation}

In the two-dimensional setting, the nature of the instability at onset is determined by the fourth-order term in the energy expansion. Computing the fourth variation yields
\begin{equation}
\begin{aligned}
E_4&=-\frac{\pi}{64}\,\frac{(m^2-1)^2}{m^4}\Bigg(
(20m^6-25m^4+2m^2-3)+\meancurv_0^2(3m^4-2m^2+1) \\
&\qquad\qquad\qquad\qquad
+ K_A(3\chi^2-1)(-3m^4+2m^2-1)
\Bigg).
\end{aligned}
\end{equation}
At the critical point $K_A=K_A^*$, where the quadratic coefficient $E_2$ vanishes, the sign of $E_4$ dictates the type of bifurcation. If $E_4>0$, the bifurcation is supercritical and stable finite-amplitude shapes emerge smoothly. If $E_4<0$, the bifurcation is subcritical, with a jump to a finite-amplitude deformation and possible hysteresis. Substituting $K_A=K_A^*$ yields
\begin{equation}
\begin{aligned}
\at{E_4}{K_A=K_A^*}
&=-\frac{\pi}{64}\,\frac{(m^2-1)^2}{m^4}\Bigg(
(20m^6-25m^4+2m^2-3)+ \meancurv_0^2(3m^4-2m^2+1)\\
&\qquad\qquad\qquad\qquad
+\frac{\meancurv_0^2+2m^2-3}{\chi^2-1}\,(3\chi^2-1)(-3m^4+2m^2-1)
\Bigg).
\end{aligned}
\end{equation}

The first nontrivial shape instability typically occurs in the $m=2$ mode. At criticality and $m=2$, the quartic coefficient reduces to
\begin{equation}
\label{eq:E4crit_m2}
% \left.E_4\right|_{K_A=K_A^*,\,m=2}
% =-\frac{9\pi}{1024}\Bigg(
% 885+41\meancurv_0^2
% -41\,\frac{(3\chi^2-1)(\meancurv_0^2+5)}{\chi^2-1}
% \Bigg).
\left.E_4\right|_{K_A=K_A^*,\,m=2} = \frac{9\pi}{512(\chi^2-1)} B(\chi,\meancurv_0), \quad B(\chi,\meancurv_0)= 340 - \chi^2(135-41\meancurv_0^2).
\end{equation}
% For convenience, define the function
% \begin{equation}
% \label{eq:Bdef}
% B(\chi,H_0)= 340 - \chi^2(135-41\meancurv_0^2).
% \end{equation}
\begin{figure}
    \centering
    \includegraphics[width=\linewidth]{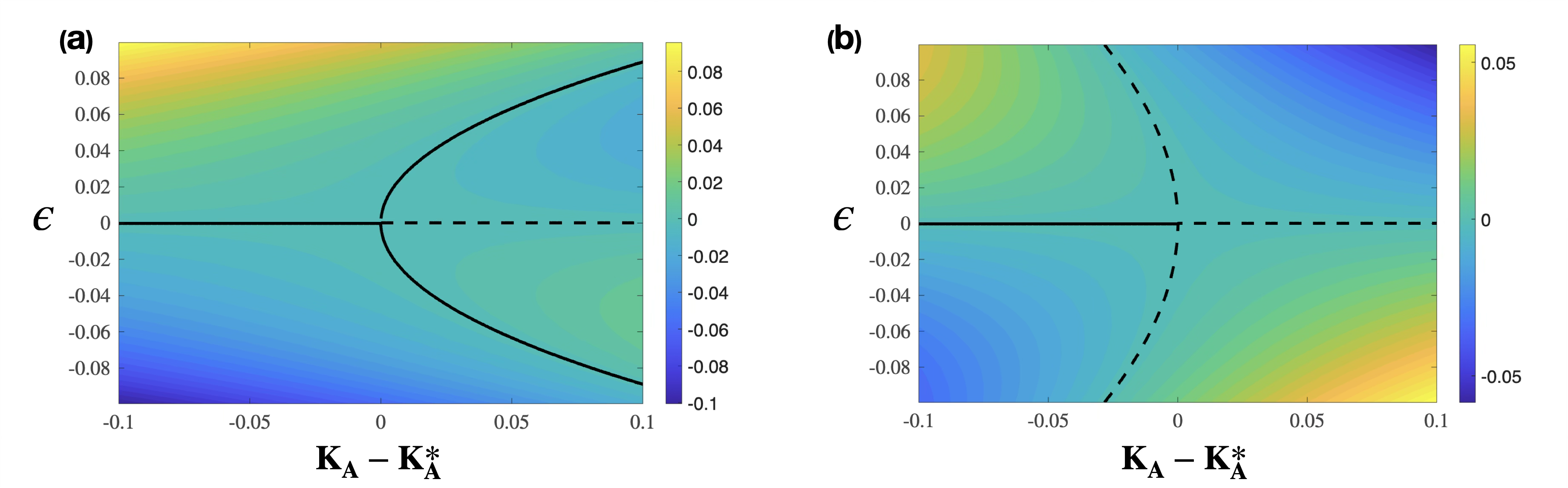}
    \caption{Bifurcation diagram for two-dimensional compressible membrane near onset of instability of circular shape. Lines indicate fixed shape $\eps$ with respect to parameter $\Areamod$. Solid line indicates stable shapes and dotted lines indicate unstable. Color indicates value of $\partial_\eps E(\eps,\Areamod)$.  (a):
    $\chi=1.3<\chi^\ast$ leading to supercritical pitchfork bifurcation. (b): $\chi=3>\chi^\ast$ leading to subcritical pitchfork bifurcation.}
    \label{fig:2dbifurcation}
\end{figure}

\begin{figure}[htbp]
    \centering
    \includegraphics[width=0.5\linewidth]{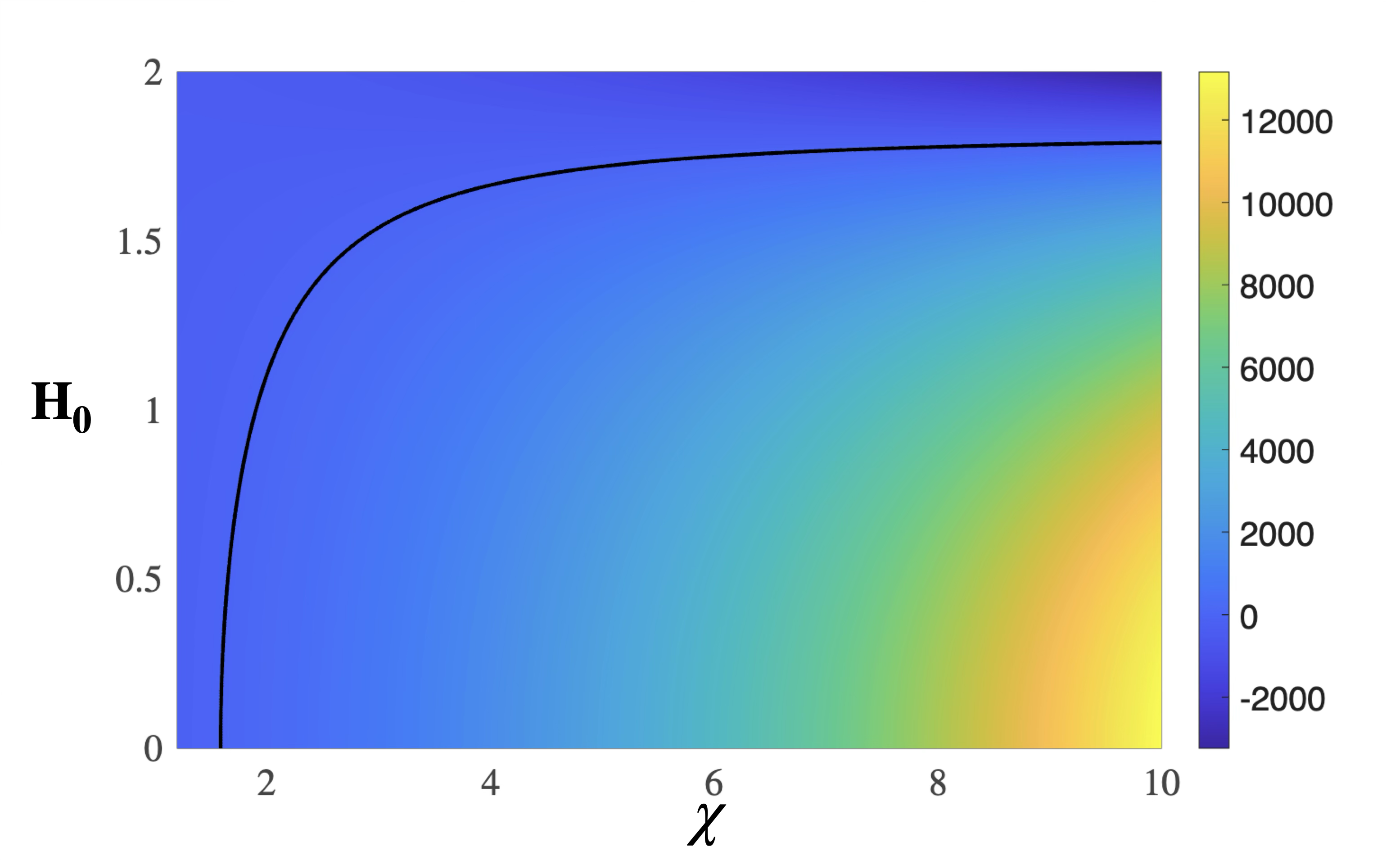}
    \caption{Diagram of the function $B(\chi,\meancurv_0)$. 
    The solid curve indicates the zero level set $B(\chi,\meancurv_0)=0$, separating parameter regimes with $B>0$ and $B<0$.
    }
    \label{fig:B_chi_H0}
\end{figure}

% Let $\Delta = K_A-K_A^*$. 

The sign of $B(\chi,\meancurv_0)$ controls the sign of $E_4$, since the remaining prefactor is positive for $\chi>1$. Thus $B>0$ implies $E_4>0$ and a supercritical bifurcation, whereas $B<0$ implies $E_4<0$ and a subcritical bifurcation; see \reffig{fig:B_chi_H0}. Assume $\meancurv_0^2<\frac{135}{41}$ and define
\begin{equation}
\label{eq:chi_c}
\chi_c^2=\frac{340}{135-41 \meancurv_0^2}\qquad(>1).
\end{equation}
Then $B(\chi, \meancurv_0)$ changes sign precisely at $\chi=\chi_c$, and
\begin{equation}
\left.E_4\right|_{K_A=K_A^*,\,m=2}
\begin{cases}
>0, & 1<\chi<\chi_c \quad\text{(supercritical)},\\[2pt]
<0, & \chi>\chi_c \quad\text{(subcritical)}.
\end{cases}
\end{equation}
Physically, increasing $\chi$ corresponds to increasing the preferred arclength relative to the fixed lipid mass, that is, a stronger energetic incentive to dilute the membrane by increasing perimeter. Larger $\chi$ therefore tends to favor a discontinuous, subcritical transition to an elongated state once bending can no longer stabilize the circle.
If instead $\meancurv_0^2\ge \frac{135}{41}$, then $B(\chi,\meancurv_0)>0$ for all $\chi>1$, and therefore
\begin{equation}
\left.E_4\right|_{K_A=K_A^*,\,m=2}>0
\qquad\text{for all }\chi>1,
\end{equation}
so the $m=2$ bifurcation is supercritical in this regime. In this case, spontaneous curvature provides an additional bending preference that regularizes the onset and yields a smooth emergence of small-amplitude prolate/oblate shapes.

\subsection{Comparison of bifurcation between two- and three- dimensions}
The stability condition of the base trivial shape in two- and three dimensions consists of the competing stabilization of the Helfrich bending and the destabilizing tension forces:
\begin{align}
    &\text{2D}: \quad K_A\paren{\chi^2-1}<\paren{2 m^2-3}+{\meancurv_0^2}\\
    &\text{3D}: \quad \Areamod\paren{\chi^2-1}<\frac{1}{2}\left[j(j+1)+2\meancurv_0(\meancurv_0+2)\right]
\end{align}
leading to similar form and asymptotic scaling for the critical $\Areamod$ values:
\begin{align}
    &\text{2D}:\quad K_A^*(m) = \frac{\meancurv_0^2+2m^2-3}{\chi^2-1}\\
    &\text{3D}:\quad \Areamod^\ast(j) = \frac{j(j+1)+2\meancurv_0(\meancurv_0+2)}{2(\chi^2-1)}.
\end{align}
The bifurcations themselves differ significantly between the two geometries. 
In three dimensions, there is a symmetry breaking between prolate $f_{2,0}>0$ and oblate $f_{2,0}<0$ shapes. 
They are geometrically distinct objects resulting in the asymmetric transcritical bifurcation near the onset of instability for the sphere.
The spontaneous curvature prescribes the `direction' or the slope of the transcritical bifurcation.
In two dimensions there is no such distinction between shapes $r=(1+f_2\cos(2\theta))$ and $r=(1-f_2\cos(2\theta))$.
One can be transformed into the other via rotation of the shape by $\pi/2$. 
Due to this symmetry, there is no odd order variation and the bifurcation in two dimensions is a pitchfork bifurcation. 
In three dimensions the slope of the transcritical bifurcation relied only on the sign of $\meancurv-\frac{3}{5}$.
The corresponding term in two dimensions which determines if the pitchfork bifurcation is super- or subcritical depends on both the spontaneous curvature $\meancurv_0$ and the baseline density $\chi$. 

\section{Conclusion}\label{sec:conclusion}
We have formulated a finite-compressibility model for a lipid-bilayer membrane in Stokes flow, replacing pointwise area incompressibility by explicit conservation of number of phospholipids. In this model, membrane tension is not a Lagrange multiplier but a constitutive response to density variation. Its gradients generate Marangoni stresses, while excess lipid density can produce an effective negative tension, providing a direct mechanism for destabilizing a bending-stabilized circular or spherical membrane.

The stability analysis shows that this mechanism is controlled by the competition between bending elasticity, which favors the base shape, and density relaxation, which favors area-increasing deformations. In both two and three dimensions, the first instability occurs in the lowest nontrivial shape mode: \(m=2\) for a circle and \(j=2\) for a sphere. The coupled membrane--Stokes dynamics also contains a fast density-dominated relaxation mode, indicating that density perturbations are not passively advected but relax through tension gradients and the hydrodynamic response of the surrounding fluid.

Near onset, the bifurcation structure depends on geometry: In two dimensions, the two signs of the \(m=2\) deformation are related by rotation, giving a pitchfork bifurcation whose criticality depends on \(\meancurv_0\) and the baseline density ratio. In three dimensions, prolate and oblate \(j=2\) perturbations are geometrically distinct, giving a generically transcritical bifurcation. The spontaneous curvature selects the preferred branch, with the transition between prolate- and oblate-favoring behavior occurring at \(\meancurv_0=3/5\), consistent with the results for the inextensible membrane under the same curvature convention.

Thus the model gives a controlled compressible extension of classical vesicle mechanics. As \(K_A\to\infty\), the density deviation vanishes and the constitutive tension approaches the incompressible Lagrange multiplier. At finite \(K_A\), the formulation retains density-driven tension, Marangoni relaxation, and lipid-density-induced shape instability. Future extensions could incorporate density-dependent material parameters, permeability, rupture, pore formation, or topological transitions.

\begin{acknowledgments}
S. K. and Y.N. Young acknowledge funding
from NSF (DMS-1951600 and DMS-2510714) and support from Flatiron Institute, part of Simons Foundation. Y. Mori acknowledges funding from NSF (DMR-2309034), and the Math+X award 234606 from the Simons Foundation.
\end{acknowledgments}

% \textcolor{black}{We have formulated a compressible membrane model that tracks phospholipid density along the interface. Unlike incompressible models, in which membrane tension appears as a Lagrange multiplier enforcing local area conservation, the present formulation gives a constitutive relation between lipid density and membrane tension. We derived stability criteria for circular membranes in two dimensions and spherical membranes in three dimensions. The energy expansion shows that the two-dimensional instability is a pitchfork bifurcation, whereas the three-dimensional instability is generically transcritical because prolate and oblate perturbations are geometrically distinct. In two dimensions, the spontaneous curvature and baseline density determine whether the pitchfork bifurcation is supercritical or subcritical. In three dimensions, the spontaneous curvature determines the slope of the transcritical branch, with a sign change at $\meancurv_0=3/5$.}

% \textcolor{black}{The model also provides a natural framework for allowing material parameters, such as bending modulus or spontaneous curvature, to depend on local phospholipid density. The effect of such density-dependent parameters on membrane response to flow remains an open direction. Because the formulation resolves density-driven tension and local area change, it may also provide a starting point for studying density-driven pathways toward rupture or open membranes, although pore formation and topology change would require additional physics such as edge energy and rupture criteria.}

\appendix

\section{Free-energy due to density variation}\label{appsec:FreeEnergy}
Let $a\in\Gamma_0$ denote a material point on a reference surface $\Gamma_0$, and let
\begin{equation}
\vb{x}=\bm X(a,t)\in \Gamma(t)
\end{equation}
be the motion of that material point. Denote by $d\Gamma_0$ the area element on the reference surface and by $d\Gamma$ the area element on the current surface. The local surface Jacobian is defined by
\begin{equation}
\Js(a,t)=\frac{d\Gamma}{d\Gamma_0},
\qquad d\Gamma=\Js\,d\Gamma_0.
\label{eq:Js_def}
\end{equation}

We use $\rho_0(a)$ to denote the reference areal lipid mass density (mass per reference area), and let $\rho(\bm X(a,t),t)$ be the current areal lipid mass density (mass per current area). For every material patch $U_0\subset\Gamma_0$, the corresponding current patch $U(t)=\bm X(U_0,t)$ contains the same lipid mass provided there is no lipid creation, destruction, or flux across the boundary of that material patch. Hence
\begin{equation}
\int_{U(t)} \rho\,d\Gamma
=\int_{U_0} \rho_0\,d\Gamma_0.
\label{eq:material_mass_patch}
\end{equation}
Since this must hold for every material patch, the local form is
%\begin{equation}
$\rho\,d\Gamma=\rho_0\,d\Gamma_0$.
%\label{eq:local_mass_exact}
%\end{equation}
Using \eqref{eq:Js_def}, one obtains the relation
\begin{equation}
\rho\,\Js=\rho_0,
\qquad\text{or equivalently}\qquad
\rho=\frac{\rho_0}{\Js}.
\label{eq:rho_J_exact}
\end{equation}
This is the connection between the density field and the local area change. It is a local statement attached to a material point, equivalent to the statement that $\rho \propto \delta h$, the local membrane thickness.

Let
\begin{equation}
f(\rho)=\frac{\KA^{\rm phys}}{2}\left(\frac{\rho}{\rhost}-1\right)^2,
\end{equation}
a near-equilibrium constitutive expansion of the free-energy density around a base state density $\rhost$.
Then the areal free energy
\begin{equation}
\label{eq:areal_free_energy}
E_\rho=\int_{\Gamma} f(\rho)\,d\Gamma.
\end{equation}
For a material deformation, local mass conservation implies
\begin{equation}
\delta(\rho\,d\Gamma)=0
\quad\Longrightarrow\quad
\delta\rho=-\rho\,\divs\vb{u}.
\label{eq:delta_rho}
\end{equation}
With $\delta(d\Gamma)=(\divs\vb{u})d\Gamma$, the variation of $E_{\rho}$ with respect to the surface deformation yields
\begin{align}
\delta E_\rho
&=\int_{\Gamma}\qty[f'(\rho)\delta\rho+f(\rho)\divs\vb{u}]\,d\Gamma \\
&=\int_{\Gamma}\qty[f(\rho)-\rho f'(\rho)]\divs\vb{u}\,d\Gamma=\int_{\Gamma}\gamma\grads\cdot\vb{u}d\Gamma = -\int_{\Gamma}\qty(\grads \gamma + 2H\gamma\vb{n})\cdot \vb{u}\,d\Gamma,
\end{align}
where the tension $\gamma\equiv\frac{\KA^{\rm phys}}{2}\left(1-\left(\frac{\rho}{\rhost}\right)^2\right)$. 
The areal free energy in (\ref{eq:areal_free_energy}) corresponds to the two-leaflet model \cite{shi2014dynamics} with a suppressed or fast relaxing leaflet-difference mode.

As a simple illustration, we use the free energy of a nearly planar bilayer membrane in Monge form
\begin{equation}
F = \int \left\{ \frac{1}{2}\kappa (\nabla^2 h)^2 + \frac{1}{2}k\left[\bigl(\rho^+ + d\nabla^2 h\bigr)^2 + \bigl(\rho^- - d\nabla^2 h\bigr)^2\right] \right\} \, dx\,dy.
\label{eq:SB12}
\end{equation}
Here $h(x,y)$ is the height field, $\kappa$ is the bare bending modulus, $k$ is a monolayer area-compressibility modulus, $d$ is the distance from the bilayer midsurface to the monolayer neutral surface, and $\rho^\pm$ are scaled density deviations of the outer and inner monolayers.
In this free energy the compressibility energy is minimized not at $\rho^\pm=0$ independently, but at the curvature-shifted values
\begin{equation}
\rho^+_{\mathrm{pref}} = -d\nabla^2 h,
\qquad
\rho^-_{\mathrm{pref}} = +d\nabla^2 h.
\label{eq:preferred_leaflet_densities}
\end{equation}
Thus bending a bilayer generates opposite preferred density changes in the two leaflets. Expanding the compressibility part of \eqref{eq:SB12} gives
\begin{align}
&\bigl(\rho^+ + d\nabla^2 h\bigr)^2 + \bigl(\rho^- - d\nabla^2 h\bigr)^2 \nonumber\\
&= (\rho^+)^2 + (\rho^-)^2 + 2d(\rho^+ - \rho^-)\nabla^2 h + 2d^2(\nabla^2 h)^2.
\label{eq:direct_expand}
\end{align}
Therefore
\begin{equation}
F = \int \left[ \frac{1}{2}(\kappa+2kd^2)(\nabla^2 h)^2 + kd(\rho^+-\rho^-)\nabla^2 h + \frac{k}{2}\bigl((\rho^+)^2+(\rho^-)^2\bigr) \right] \, dx\,dy.
\label{eq:expanded_leaflet_form}
\end{equation}
% The effective bending modulus in this form is
% \begin{equation}
% \kappa_e = \kappa + 2kd^2.
% \label{eq:kappa_eff}
% \end{equation}
%
Next we introduce
\begin{equation}
\dbar\rho = \frac{\rho^+ + \rho^-}{2},
\qquad
\Delta\rho = \frac{\rho^+ - \rho^-}{2}.
\label{eq:sym_antisym_def}
\end{equation}
Then
\begin{equation}
\rho^+ = \bar\rho + \Delta\rho,
\qquad
\rho^- = \bar\rho - \Delta\rho,
\label{eq:invert_sym_antisym}
\end{equation}
and hence
\begin{equation}
(\rho^+)^2+(\rho^-)^2 = 2\bar\rho^2 + 2\Delta\rho^2,
\qquad
\rho^+-\rho^- = 2\Delta\rho.
\label{eq:identities_sym_antisym}
\end{equation}
% Substitution into \eqref{eq:expanded_leaflet_form} yields
% \begin{equation}
% F = \int \left[ \frac{1}{2}\kappa_e(\nabla^2 h)^2 + 2kd\,\Delta\rho\,\nabla^2 h + k\bar\rho^2 + k\Delta\rho^2 \right] \, dx\,dy.
% \label{eq:symantisym_form}
% \end{equation}
%
We identify variable $\bar\rho$ as the common-mode density deviation: both leaflets are simultaneously compressed or diluted relative to equilibrium.
The variable $\Delta\rho$ is the leaflet-difference mode: one leaflet is compressed while the other is diluted. This is the degree of freedom that carries interleaflet mismatch and couples directly to curvature.
%
% Thus the bilayer theory contains two density sectors:
% \begin{equation}
% F_{\mathrm{dens}} = \int \bigl(k\bar\rho^2 + k\Delta\rho^2\bigr)\,dxdy,
% \label{eq:dens_sector_only}
% \end{equation}
% plus the coupling term
% \begin{equation}
% F_{\mathrm{couple}} = \int 2kd\,\Delta\rho\,\nabla^2 h\,dxdy.
% \label{eq:coupling_sector_only}
% \end{equation}

The terms involving $\Delta\rho$ in the free energy density may be recast as
\begin{align}
\frac{1}{2}(\kappa+2kd^2)(\nabla^2 h)^2 + 2kd\,\Delta\rho\,\nabla^2 h + k\Delta\rho^2 
&= \frac{1}{2}\kappa(\nabla^2 h)^2 + k\bigl(\Delta\rho + d\nabla^2 h\bigr)^2.
\label{eq:complete_square}
\end{align}
Hence
\begin{equation}
F = \int \left[ \frac{1}{2}\kappa(\nabla^2 h)^2 + k\bar\rho^2 + k\bigl(\Delta\rho + d\nabla^2 h\bigr)^2 \right] \, dx\,dy.
\label{eq:completed_square_final}
\end{equation}
This makes the preferred antisymmetric density explicit:
\begin{equation}
\Delta\rho_* = -d\nabla^2 h.
\label{eq:Delta_opt}
\end{equation}
If the leaflet-difference mode relaxes rapidly, then it is slaved to curvature by \eqref{eq:Delta_opt}, and the reduced free energy becomes
\begin{equation}
F_{\mathrm{red}} = \int \left[ \frac{1}{2}\kappa(\nabla^2 h)^2 + k\dbar\rho^2 \right] \, dx\,dy.
\label{eq:F_reduced_eliminate_Delta}
\end{equation}

Next we let the absolute areal densities of the outer and inner leaflets be denoted by $\rho^+_{\mathrm{abs}}$ and $\rho^-_{\mathrm{abs}}$, respectively.
Let their common equilibrium value in a flat reference configuration be $\rho^*_{\mathrm{mono}}$.
Define the dimensionless leaflet density deviations by
\begin{equation}
\rho^+ = \frac{\rho^+_{\mathrm{abs}}-\rho^*_{\mathrm{mono}}}{\rho^*_{\mathrm{mono}}},
\qquad
\rho^- = \frac{\rho^-_{\mathrm{abs}}-\rho^*_{\mathrm{mono}}}{\rho^*_{\mathrm{mono}}}.
\label{eq:dimensionless_leaflet_devs}
\end{equation}
This is the natural interpretation of the leaflet variables appearing in \eqref{eq:SB12}.
Now define the effective bilayer areal density as the average of the two absolute leaflet densities:
\begin{equation}
\rho_{\mathrm{eff}} = \frac{\rho^+_{\mathrm{abs}}+\rho^-_{\mathrm{abs}}}{2}.
\label{eq:rho_eff_def}
\end{equation}
Its equilibrium value is
\begin{equation}
\rho^*_{\mathrm{eff}} = \rho^*_{\mathrm{mono}}.
\label{eq:rho_eff_star}
\end{equation}
Subtracting \eqref{eq:rho_eff_star} from \eqref{eq:rho_eff_def} gives
\begin{align}
\rho_{\mathrm{eff}} - \rho^*_{\mathrm{eff}}
&= \frac{1}{2}\left[(\rho^+_{\mathrm{abs}}-\rho^*_{\mathrm{mono}}) + (\rho^-_{\mathrm{abs}}-\rho^*_{\mathrm{mono}})\right] \nonumber\\
&= \frac{\rho^*_{\mathrm{mono}}}{2}(\rho^+ + \rho^-) \nonumber\\
&= \rho^*_{\mathrm{mono}}\,\dbar\rho.
\label{eq:careful_map}
\end{align}
This shows that the free energy in (\ref{eq:F_reduced_eliminate_Delta}) is a bending-plus-density-penalty theory.

Remark: The common-mode leaflet deviation $\dbar\rho$ is proportional to the deviation of the effective bilayer areal density from its preferred value. Therefore $\dbar\rho$ is the natural bilayer quantity that corresponds to the one-field density deviation, but only after one passes through the absolute-density definition \eqref{eq:rho_eff_def}. 
% The correct statement is therefore not
% \begin{equation*}
% \rho = \rho^* + \bar\rho,
% \end{equation*}
% but rather
% \begin{equation}
% \rho_{\mathrm{eff}} - \rho^*_{\mathrm{eff}} = \rho^*_{\mathrm{mono}}\,\bar\rho.
% \label{eq:correct_map_statement}
% \end{equation}
If the one-field theory uses a dimensionless density deviation instead of an absolute density deviation, then one may divide by $\rho^*_{\mathrm{eff}}$ and obtain $\dbar\rho=\frac{\rho_{\mathrm{eff}}-\rho^*_{\mathrm{eff}}}{\rho^*_{\mathrm{eff}}}$.
In that case the common-mode bilayer formulation is exactly the dimensionless one-field density framework.

\section{Spherical Harmonics and Stokes Basis}
\label{app:sphericalharmonics}

\subsection{Spherical Harmonics}

The notation convention and basis match those given in \cite{vlahovska2015dynamics}. The normalized scalar spherical harmonic are defined as 
\begin{equation}
\Yspp_{jm}(\theta,\phi) = \bigg[\frac{2j+1}{4\pi} \frac{(j-m)!}{(j+m)!}\bigg]^{1/2} P_{j}^{m}(\cos(\theta))e^{i m \phi}\label{def:sphericalharmonics}
\end{equation}
where $\theta$ and $\phi$ are the polar and azimuthal angles in spherical coordinates, and $P_{j}^m$ are the associated Legendre polynomials.

The vector spherical harmonics are defined as
\begin{equation}
\begin{split}
&\by_{jm0} = \frac{1}{\sqrt{j(j+1)}}r\bnab _{\Omega} Y_{jm},\,\quad\by_{jm1} = -i \rhat \times \by_{jm0},\,\quad\by_{jm2} = Y_{jm} \rhat
\end{split}\label{def:vsphericalharmonics}
\end{equation}
where $\bnab _{\Omega} $ denotes the angular part of the gradient operator. 

We have the following identities
\begin{equation}
\begin{split}
    &\bnab_\Omega\times \by_{j,m,0} = \bm{0}, \quad \bnab_\Omega\times \by_{j,m,1} =i\by_{j,m,0}+i\sqrt{j(j+1)}\by_{j,m,2},\quad \bnab_\Omega\times \by_{j,m,2} = -i\sqrt{j(j+1)}\by_{j,m,1}\\
    &\bnab_\Omega\times\bnab_\Omega\times \by_{j,m,0} =\bm{0}, \quad \bnab_\Omega\times\bnab_\Omega\times \by_{j,m,1} =j(j+1)\by_{j,m,1}\\
    &\bnab_\Omega\times\bnab_\Omega\times \by_{j,m,2} = \sqrt{j(j+1)}\by_{j,m,0}+j(j+1)\by_{j,m,2}
    \end{split}
\end{equation}

\subsection{Fundamental set of solutions for the Stokes equation in three dimensions}\label{app:fundamentalflows}
Following the definitions in \cite{vlahovska2015dynamics}, we list a  basis for solutions to the Stokes equations:
\begin{equation}
\begin{split}
&\bu^-_{jm0} = \frac{1}{2}r^{-j}(2-j+jr^{-2})\by_{jm0}+\frac{1}{2}r^{-j}\sqrt{j(j+1)}(1-r^{-2})\by_{jm2}\\
&\bu_{jm1}^- = r^{(-j-1)}\by_{jm1}\\
&\bu_{jm2}^- = \frac{1}{2}r^{-j}(2-j)\sqrt{\frac{j}{j+1}}(1-r^{-2})\by_{jm0}+\frac{1}{2}r^{-j}(j+(2-j)r^{-2})\by_{jm2}\\
&\bu_{jm0}^+ = \frac{1}{2}r^{j-1}(-(j+1)+(j+3)r^2)\by_{jm0}-\frac{1}{2}r^{j-1}\sqrt{j(j+1)}(1-r^2)\by_{jm2}\\
&\bu_{jm1}^+ = r^j\by_{jm1}\\
&\bu_{jm2}^+ = \frac{1}{2}r^{j-1}(j+3)\sqrt{\frac{j+1}{j}}(1-r^2)\by_{jm0}+\frac{1}{2}r^{j-1}(j+3-(j+1)r^2)\by_{jm2}.
\end{split}\label{eqapp:3Dstokesbasis}
\end{equation}
On the unit sphere, the velocity fields reduce to
\begin{equation}
\bv^{\pm}_{jm\q} = \by_{jm\q}.
\end{equation}
Each of the basis fields are also incompressible everywhere except possibly at the origin.
% \begin{equation}
% \bnab \cdot\bv^{\pm}_{jm\q} = 0.
% \end{equation}
%\subsection{Hydrodynamic stresses}
Given a velocity of the following form
\begin{equation}
\bv = \sumjmq c_{jm\q}^\pm \bu_{jm\q}^{\pm},
\end{equation}
The $T_{jm\q}^\pm$ used in bulk hydrodynamic stress are defined via the radial traction of $\bu$:
\begin{equation}
\rhat \cdot (-p\bm{I}+\nabla \bu+(\nabla \bu)^T) = \sum_{j,m,s}\tau_{jm\q}^{\HD,\pm}\by_{jm\q},\quad\tau_{jm\q}^{\HD,\pm} = \sum_{\q=0}^2 c_{jm\q'}^\pm T^\pm_{\q\q'}\label{eqapp:HDstress}
\end{equation}
where

\begin{equation}
T_{\q\q'}^-=\begin{bmatrix}
-(2j+1)&0&3\sqrt{\frac{j}{j+1}}\\
0&-(j+2)&0\\
 3\sqrt{\frac{j}{j+1}}&0&-\frac{4+3j+2j^2}{j+1}
\end{bmatrix},\quad T_{\q\q'}^+=\begin{bmatrix}
 (2j+1)&0&-3\sqrt{\frac{j+1}{j}}\\
0&(j-1)&0\\
  -3\sqrt{\frac{j+1}{j}}&0&\frac{3+j+2j^2}{j}
\end{bmatrix}.
\end{equation}

\section{Flow solution for three-dimensional problem}\label{app:flowsolution}
\subsection{Interface parametrization}\label{appsub:interfaceparametrization}
The shape and density of the membrane are parametrized in \refeqn{eq:3Dparametrization}, and \refeqn{eq:3Dparaverage} as
\begin{equation}
    \begin{split}
        &\bx_\surf(\theta,\phi) = \big(R+\eps f(\theta,\phi)\big)\rhat, \quad \rho(\theta,\phi) = \rho_0+ \eps g(\theta,\phi)  \\
        &R = 1 -\frac{1}{4\pi}\sum_{n=2}^\infty \eps^n \Delta_{V,n},\quad \rho_0 = \chi-\frac{1}{4\pi}\sum_{n=2}^\infty\eps^n\Delta_{\rho,n}
    \end{split}
\end{equation}

% A nearly spherical membrane is parametrized as
% \begin{equation}
%     \bx_\surf = \big(R+\eps f(\theta,\phi)\big)\rhat
%     \label{eq:3Dshapeparametrization2}
% \end{equation}
% where the average radius $R$ is chosen to preserve the volume of the enclosure under deformation:
% \begin{equation}
%     R = 1 -\frac{1}{4\pi}\sum_{n=2}^\infty \eps^n \Delta_{V,n}. \label{eq:3Daverad2}
% \end{equation}
The first two non-zero corrections of order $\eps^2$ and $\eps^3$ in \refeqn{eq:3Dparaverage} to the average radius and average density are given by
\begin{equation}
\begin{split}
    &\Delta_{V,2} = \int_{\surf_0} f^2 \,d\surf_0,\quad\Delta_{V,3} = \frac{1}{3}\int_{\surf_0} f^3\,d\surf_0\\
    &\Delta_{\rho,2} = \int_{\surf_0} -f^2+\frac{1}{2}(\bnab_0f)\cdot(\bnab_0 f)+2fg\, \,d\surf_0\\
    &\Delta_{\rho,3} = \int_{\surf_0} 2f^3-\frac{1}{2}g(2f+(\bnab_0f)\cdot(\bnab_0f)\,d\surf_0 
    \end{split}\label{eqapp:3Davgcorrections}
\end{equation}
where$\bnab_0$ is the surface gradient over a unit sphere and $\Gamma_0$ is integration over the unit sphere.

\subsection{Interfacial   stresses}\label{subapp:interfacialstresses}
Given parametrization \refeqn{eq:3Dparametrization} for the surface and density, the non-isotropic component of the Helfrich bending stress and stress from density driven tension at leading order in $\eps$
\begin{equation}
    \tau^B = \sum_{j,m}\tau_{j,m,\q}^B\Yspp_{j,m,\q} , \quad \tau^E = \sum_{j,m}\tau_{j,m,\q}^E\Yspp_{j,m,\q}\label{eqapp:interfacialstresses}    
\end{equation}
where
\begin{equation}
    \begin{split}
        &\tau^B_{j,m,2} = \eps\frac{\Bendmod}{2}(j-1)(j+2)\big(2\meancurv_0(\meancurv_0+2)+j(j+1)\big)\fjm\\
        &\tau^E_{j,m,0} = \Areamod\eps\sqrt{j(j+1)}\gjm\\
        &\tau^E_{j,m,2} = \Areamod\eps\big(-2(\chi-1)(j-1)(j+2)\fjm-2\gjm\big)\Yjmtwo 
\end{split}\label{eqapp:3Dstressessph}
\end{equation}
where the tension has been linearized with respect to $(\chi-1)\ll 1$.
The isotropic component of the stress has been removed as it does not contribute to the dynamics.

\subsection{Flow solution}
% Given parametrization \refeqn{eq:3Dparametrization} for the surface and density, the non-isotropic component of the Helfrich bending stress and stress from density driven tension at leading order in $\eps$
% \begin{equation}
%     \begin{split}
%         &\tau^B = \eps\frac{\Bendmod}{2}(j-1)(j+2)\big(2H_0(H_0+2)+j(j+1)\big)\fjm\Yjmtwo\\
%         &\tau^E = \Areamode\eps\bigg[\big(-2(\rhobar-1)(j-1)(j+2)\fjm-2\gjm\big)\Yjmtwo +\sqrt{j(j+1)}\gjm\Yjmzero\bigg]
% \end{split}\label{eq:3Dstressessph}
% \end{equation}
% where it has been assumed that $(\rhobar-1)\sim O(1)$.
% The isotropic components of the stress, leading to constant pressure solutions inside and outside the membrane, do not drive flow or shape deformation and have been removed.
% The $\Yjmtwo$ component of the stresses are normal to the surface of a unit sphere and govern the shape dynamics for nearly spherical shapes.
% The $\Yjmzero$ components are tangential to the unit sphere and drive redistribution of the lipid density on the interface.

The flow inside and outside the membrane is given in \refeqn{eq:3Dsolansatz}:
\begin{equation}
    \bu^\out = \sumjmq c_{j,m,\q}\bu_{j,m,\q}^-, \quad\bu^\ins = \sumjmq c_{j,m,q}\bu_{j,m,\q}^+. 
\end{equation}
The stress balance conditions, \refeqn{eq:3DBC3}, \refeqn{eq:stressbalancesph} leads to solution
% The leading order hydrodynamic stress is given by 
% \begin{equation}
%     \tau^{\HD} = \sumjmq \tau_{j,m,\q}^{\HD,\pm} \by_{j,m,\q}.
% \end{equation}
% where $\tau_{jm\q}^{\HD,\pm} = \sum_{\q'=0}^2 c_{jm\q'}^\pm T^\pm_{\q\q'}$ and $T^\pm_{\q\q'}$ are given in \refapp{app:sphericalharmonics}. 
% Solving the resulting equation $\tau^{\HD} = \tau^B+\tau^E$ gives
\begin{equation}
    \begin{split}
    &c_{j,m,0} = \eps\big(\sqrt{ j(j+1)}\xi_{j,m}\big)^{-1}\bigg[ p_{j,m,0}^f\fjm+p_{j,m,0}^g\gjm\bigg] \\
    &c_{j,m,1} =0 ,\quad
    c_{j,m,2} = \eps\xi^{-1}_{j,m}\bigg[ p_{j,m,2}^f\fjm+p_{j,m,2}^g\gjm\bigg] 
    \end{split}\label{eqapp:3Dcoefsol}
\end{equation}
where
\begin{equation}
\begin{split}
    &\xi = (2j+1)(2j-1)(2j+3)\\
    &p_{j,m,0}^f = \frac{3}{2}(j-1)j(j+1)(j+2)(2j+1)\big(2\Areamod(\chi-1)-(2\meancurv_0(\meancurv_0+2)+j(j+1))\big)\\
    &p_{j,m,2}^f = j(j-1)(j+1)(j+2)(2j+1)(2\Areamod(\chi-1)-\big(2\meancurv_0(\meancurv_0+2)+j(j+1)\big)\big)\\
    &p_{j,m,0}^g=-\Areamod j(j+1)(2j^2+2j-3),\quad p_{j,m,2}^g = \Areamod j(j+1)
    \end{split}.\label{eqapp:3Dcoefsol2}
\end{equation}

%\section{Flow solution with surface viscosity}

\section{Circular membrane}\label{app:2D}
\subsection{Energy identity and interfacial condition}\label{app:2Dmodels}
% Let $\varrho(\theta,t):\mathbb S\times [0,T]\to \R$ be the lipid density at a point $\bm q(\theta,t)\in\Gamma(t)$. For any point $\bm X\in\Gamma$ at a fixed time, the density at the point is defined as $\rho(\bm q) =\varrho\circ\bm X^{-1}(\bm q)$.
% The mass in a small arc-length element is given by $\varrho(\theta,t)\abs{\partial_\theta\bm X(\theta,t)}d\theta$.
% With no dissipation, this material mass is constant along the trajectory of each Lagrangian coordinate $\theta$, that is 
% \begin{equation}
%     \frac{d}{dt}\paren{\varrho(\theta,t)\abs{\partial_\theta\bm X(\theta,t)}} = 0.
% \end{equation}
% Given the initial condition $\varrho(\theta,0) = \rhobar(\theta)/\abs{\partial_\theta\bm X(\theta,0)}$ where $\rhobar(\theta)$ is a constant in time, representing the mass distribution, we have
% \begin{equation}
%     \varrho(\theta,t) = \frac{\rhobar(\theta)}{\abs{\partial_\theta\bm X(\theta,t)}}.
% \end{equation}
% Assuming the curve is uniformly stretched (mass distribution is uniform) at $t=0$, we can set $\rhobar(\theta)$ as a constant in space as well. Hence
% \begin{equation}
%     \varrho(\theta,t) = \frac{\chi}{\abs{\partial_\theta\bm X(\theta,t)}}.
% \end{equation}

% Let $H$ be the curvature of $\Gamma$.

In the circular membrane model, the original total energy is
\begin{equation}
    E[\Gamma] = \frac{1}{2} \int_\Gamma \kappa\paren{\meancurv-\meancurv_0}^2 + K_A^{\rm phys}\paren{\frac{\rho}{\rho^*}-1}^2\,ds,
\end{equation}
where $\kappa$, $\meancurv$, $\meancurv_0$, $K_A^{\rm phys}$ and $\rho^*$ are the bending modulus, curvature, spontaneous curvature, dimensional area-compression modulus and preferred local density, respectively.
As in Section \ref{secsub:rescaled}, the corresponding nondimensional Stokes system is
\begin{equation}\label{eq:2Ddynamicalequation}
\begin{aligned}
    -\Delta\bm{u}+\nabla p=&0,\\
    \nabla\cdot\bm{u}=&0,\\
    \jump{\bm{u}}=&0,\\
    \jump{\bsigma(\bm u,p)\bm n} =& \bm F,\\
    \partial_t\bm X=&\bm{u}(\bm X).
\end{aligned}    
\end{equation}
% consider the bending rigidity is rescaled to one.
% and $K_A$ is the nondimensional area expansion modulus.
and the nondimensional energy is
\begin{equation}
\begin{aligned}
    E[\Gamma] = \frac{1}{2} \int_\Gamma \paren{\meancurv-\meancurv_0}^2 + K_A\paren{\rho-1}^2\,ds 
\end{aligned}
\end{equation}
where $\bm{F}=-\frac{1}{Q}\frac{\delta E}{\delta \bm{X}}$.
Note that $\int_\Gamma \meancurv \,ds =2 \pi$.
Then, the energy can be transformed as
\begin{equation}\label{eq:2Dtotalenergy}
\begin{aligned}
    E[\Gamma] &= \frac{1}{2} \int_\Gamma \paren{\meancurv-\meancurv_0}^2 + K_A\paren{\rho-1}^2\,ds \\
    &= \frac{1}{2} \int_\Gamma \paren{\meancurv^2 + \meancurv_0^2} + K_A\paren{\rho-1}^2\,ds -2\pi \meancurv_0\\
    &= \frac{1}{2}\int_{\mathbb S} \paren{\paren{\frac{P}{Q^3}}^2 + \meancurv_0^2+ K_A\paren{\frac{\chi}{Q} - 1}^2}Q\,d\theta- 2\pi \meancurv_0 \\
    &= \frac{1}{2}\int_{\mathbb S} \frac{P^2}{Q^5} + \meancurv_0^2Q + 
    K_A \frac{ \paren{Q - \chi}^2  }{Q} \,d\theta- 2\pi \meancurv_0,
\end{aligned}
\end{equation}
where $\chi$ is the nondimensional conserved lipid mass per unit reference arclength,
\begin{equation}
\begin{aligned}
    P\paren{\Xb}=R\pd{\theta}\Xb\cdot\pdd{\theta}{2}\Xb,\quad
    Q\paren{\Xb}=\abs{\pd{\theta}\Xb}.
\end{aligned}
\end{equation} 
and the rotation matrix $R$ is
\begin{equation}
    R = 
    \begin{bmatrix}
        0 & 1 \\
        -1 & 0
    \end{bmatrix}.
\end{equation}
We assume that the enclosed area is fixed as $\pi$ and $\chi > 1$.
Denote
% \begin{equation}
%     E[\Gamma]=E_1[\Gamma]+\meancurv_0^2~ E_2[\Gamma]+K_A~  E_3[\Gamma]-2\pi \meancurv_0
% \end{equation}
% where
\begin{equation}
\begin{aligned}
    E_1[\Gamma]=\frac{1}{2}\int_\mbs \frac{P^2}{Q^5}d\theta,~~
    E_2[\Gamma]=\frac{1}{2}\int_\mbs Q d\theta,~~
    E_3[\Gamma]=\frac{1}{2}\int_\mbs \frac{\paren{Q-\chi}^2}{Q}d\theta
\end{aligned}
\end{equation}
% and

% \begin{equation}
% \begin{aligned}
%     P\paren{\Xb}=R\pd{\theta}\Xb\cdot\pdd{\theta}{2}\Xb,\quad
%     Q\paren{\Xb}=\abs{\pd{\theta}\Xb}.
% \end{aligned}
% \end{equation}
The interfacial condition is
\begin{equation*}
    \jump{\bm \sigma(\bm u, p)\bm n} = \bm F=-\frac{1}{Q}\paren{\frac{\delta E_1}{\delta \Xb}+\meancurv^2_0\frac{\delta E_2}{\delta \Xb}+K_A\frac{\delta E_3}{\delta \Xb}}
\end{equation*}
where
\begin{equation}
\begin{aligned}
    \frac{\delta E_1}{\delta \Xb}&=\pd{\theta}\paren{\frac{P}{Q^5} R\pdd{\theta}{2}\Xb+\frac{5P^2}{2Q^7}{\pd{\theta}\Xb}}+\pdd{\theta}{2}\paren{\frac{P}{Q^5} R\pd{\theta}\Xb}\\
    \frac{\delta E_2}{\delta \Xb}&=-\pd{\theta}\paren{\frac{1}{2Q}{\pd{\theta}\Xb}}\\
    \frac{\delta E_3}{\delta \Xb}&=\pd{\theta}\paren{\frac{1}{2Q}\paren{\frac{\chi^2}{Q^2}-1}{\pd{\theta}\Xb}}
\end{aligned}
\end{equation}

\subsection{Linearization}

We first consider the steady state equation.
Setting $\partial_t\bm X = \bm u(\bm X(\theta)) = 0$ leads to
\begin{equation}
    \bm u(\bm x)\equiv \bm 0,\quad p =
    \begin{cases}
        p_i, \quad \bm x \in \Omega_i,\\
        p_e, \quad \bm x \in \Omega_e,
    \end{cases}
\end{equation}
where $p_{i,e}$ are constants.
Due to the divergence-free condition, the enclosed area is fixed, which we set as $A = \pi$.
Together with the interface condition, we have the steady state equation
\begin{align}
    &-\paren{\frac{\delta E_1}{\delta \Xb}+\meancurv^2_0\frac{\delta E_2}{\delta \Xb}+K_A\frac{\delta E_3}{\delta \Xb}}+\jump{p_0}R\partial_\theta \bm X = 0,\\
    &\frac{1}{2} \int_{\mathbb S} \bm X(\theta)\cdot R \partial_\theta\bm X(\theta) \,d\theta -\pi = 0,
\end{align}
where $\jump{p_0} = p_i-p_e$ is a constant.
Let
\begin{equation}
    \bm e_r = (\cos \theta, \sin\theta)^T,\quad \bm e_\theta = (-\sin\theta, \cos\theta)^T.
\end{equation}
There could be multiple solutions depending on the values $K_A$ and $\chi$.
Obviously, since $-P\paren{\ebr}=Q\paren{\ebr}=1$, the unit circle $\bm X = \bm e_r$ is a solution with
\begin{equation}
    \jump{p_0} = \frac{-1 + \meancurv_0^2+ K_A\paren{1-\chi^2}}{2}.
\end{equation}

Now we linearize the problem around the circular shaped steady state. Let
\begin{equation}
    \bm u = \varepsilon \bm v,\quad p = p_0 + \varepsilon q, \quad \bm X = \bm e_r + \varepsilon\bm Y,
\end{equation}
so
\begin{equation}
\begin{aligned}    
    P_\varepsilon
    =-1+\varepsilon\paren{\pdd{\theta}{2}\Yb-R\pd{\theta}\Yb}\cdot\ebr+O\paren{\varepsilon^2},~~
    \frac{1}{Q_\varepsilon}   
    =1-\varepsilon\pd{\theta}\Yb\cdot\ebtheta+O\paren{\varepsilon^2}
\end{aligned}
\end{equation}
Denote $\bm{F}_{i1}$ by
\begin{equation}
    \frac{\delta E_i}{\delta \Xb}=-\bm{F}_{i0}-\varepsilon\bm{F}_{i1}-\varepsilon^2\bm{F}_{i2}+O\paren{\varepsilon^3}
\end{equation}
Gathering the first-order terms, the linearized problem is given by
\begin{subequations}
    \begin{align}
        -\Delta \bm v + \grad q & = 0,\\
        \grad\cdot \bm v &= 0,\\
        \jump{\bm v} & = 0, \\
        \jump{\bm\sigma(\bm v, q)\bm e_r} &= \jump{p_0}R\partial_\theta\bm Y  + \bm{F}_{11}+ \meancurv_0^2 \bm{F}_{21}+K_A \bm{F}_{31},\\
        \partial_t\bm Y &= \bm v(\bm e_r).
    \end{align}
\end{subequations}
where
\begin{equation}
\begin{aligned}
    \bm{F}_{11}
    =&\pd{\theta}\left[ 2R\pdd{\theta}{2}\Yb-\frac{5}{2}\pd{\theta}\Yb+\pd{\theta}\paren{{\pdd{\theta}{2}\Yb}\cdot\ebr-4{\pd{\theta}\Yb\cdot\ebtheta}}\ebr\right.\\
     &+\left.\paren{{3\pdd{\theta}{2}\Yb}\cdot\ebr+\frac{9}{2}{\pd{\theta}\Yb\cdot\ebtheta}}\ebtheta\right]\\
     \bm{F}_{21}
     =&\pd{\theta}\left[\frac{1}{2}\pd{\theta}\Yb-\frac{1}{2}\paren{{\pd{\theta}\Yb\cdot\ebtheta}}\ebtheta\right]\\
     \bm{F}_{31}
     =&\pd{\theta}\left[\frac{1-\chi^2}{2}\pd{\theta}\Yb-\frac{1-3\chi^2}{2}\paren{{\pd{\theta}\Yb\cdot\ebtheta}}\ebtheta\right]
\end{aligned}
\end{equation}

Denote $L_{ir},L_{i\theta}$ by
\begin{equation}
    \bm{F}_{i1}=\pd{\theta}\paren{L_{ir}\paren{\Yb}\ebr}+\pd{\theta}\paren{L_{i\theta}\paren{\Yb}\ebtheta}
\end{equation}

Since
\begin{equation}
    R\pdd{\theta}{2}\Yb=-\paren{\pdd{\theta}{2}\Yb\cdot\ebr}\ebtheta+\paren{\pdd{\theta}{2}\Yb\cdot\ebtheta}\ebr,
\end{equation}

\begin{equation}
\begin{aligned}
    L_{1r}=&\paren{-\pdd{\theta}{3}\Yb+\frac{3}{2}\pd{\theta}\Yb}\cdot\ebr-3\pdd{\theta}{2}\Yb\cdot\ebtheta,~~
    L_{2r}=0,~~
    L_{3r}={\frac{1-\chi^2}{2}\pd{\theta}\Yb}\cdot\ebr\\
    L_{1\theta}=&\pdd{\theta}{2}\Yb\cdot\ebr+2\pd{\theta}\Yb\cdot\ebtheta,~~    
    L_{2\theta}=\frac{1}{2}\pd{\theta}\Yb\cdot\ebtheta,~~
    L_{3\theta}=\chi^2\pd{\theta}\Yb\cdot\ebtheta,~~
\end{aligned}
\end{equation}

% \begin{equation}
% \begin{aligned}
%     L_{2r}=&{-\frac{1}{2}\pd{\theta}\Yb}\cdot\ebr\\
%     L_{2\theta}=&-\pd{\theta}\Yb\cdot\ebtheta
% \end{aligned}
% \end{equation}

% \begin{equation}
% \begin{aligned}
%     L_{3r}=&{\frac{1-\chi^2}{2}\pd{\theta}\Yb}\cdot\ebr\\
%     L_{3\theta}=&\chi^2\pd{\theta}\Yb\cdot\ebtheta
% \end{aligned}
% \end{equation}

\subsection{Spectrum calculations}
We consider the linearized problem in the polar coordinate.
We write $\bm v$ and $\bm Y$ as
\begin{equation}
\begin{aligned}
    \bm v = v^r(r,\theta) \bm e_r + v^\theta(r,\theta) \bm e_\theta ,\quad
    \bm Y = Y^r(\theta) \bm e_r + Y^\theta(\theta) \bm e_\theta.
\end{aligned}
\end{equation}
The Stokes equation in polar coordinates is given by
\begin{align}
-\frac{\partial q}{\partial r} + \left( \frac{\partial^2 v^r}{\partial r^2} + \frac{1}{r} \frac{\partial v^r}{\partial r} - \frac{v^r}{r^2} + \frac{1}{r^2} \frac{\partial^2 v^r}{\partial \theta^2} - \frac{2}{r^2} \frac{\partial  v^\theta }{\partial \theta} \right) &= 0, \label{eqn:st-1}\\
-\frac{1}{r} \frac{\partial q}{\partial \theta} + \left( \frac{\partial^2  v^\theta }{\partial r^2} + \frac{1}{r} \frac{\partial  v^\theta }{\partial r} - \frac{ v^\theta }{r^2} + \frac{1}{r^2} \frac{\partial^2  v^\theta }{\partial \theta^2} + \frac{2}{r^2} \frac{\partial v^r}{\partial \theta} \right) &= 0,\label{eqn:st-2}
\\
 \frac{\partial}{\partial r} (r v^r) + \frac{\partial  v^\theta }{\partial \theta} &= 0.\label{eqn:st-3}
\end{align}
The interface condition at $r=1$ becomes
\begin{align}
    \jump{v^r} = \jump{v^\theta}= 0,
\end{align}
Since
\begin{equation}
    \jump{\bm\sigma(\bm v, q)\bm e_r}=\pd{\theta}\paren{L_r\paren{\Yb}\ebr}+\pd{\theta}\paren{L_\theta\paren{\Yb}\ebtheta},
\end{equation}
where
\begin{equation}
\begin{aligned}
    L_r=& L_{1r}+ \meancurv_0^2 L_{2r}+K_A L_{3r}+\jump{p_0}\Yb\cdot\ebtheta\\
    L_\theta=&L_{1\theta}+ \meancurv_0^2 L_{2\theta}+K_A L_{3\theta}-\jump{p_0}\Yb\cdot\ebr,
\end{aligned}
\end{equation}
we obtain
\begin{align}
    -\jump{q} &= \jump{\bm\sigma(\bm v, q)\bm e_r} \cdot \bm e_r=\pd{\theta}L_r-L_\theta\\
    \jump{\partial_r v^\theta} &= \jump{\bm\sigma(\bm v, q)\bm e_r} \cdot\bm e_\theta=\pd{\theta}L_\theta+L_r
\end{align}
We express the solution as Fourier series, 
\begin{equation}
\begin{aligned}
    \bm v &= \sum_{m=-\infty}^\infty \paren{ v_m^r(r) \bm e_r + v_m^\theta(r) \bm e_\theta }e^{im\theta},\\
    q &= \sum_{m=-\infty}^\infty q_m(r)e^{im\theta},\\
    \bm Y &= \sum_{m=-\infty}^\infty \paren{ Y_m^r \bm e_r + Y_m^\theta \bm e_\theta }e^{im\theta}.
\end{aligned}
\end{equation}
Since $\bm v, q, \bm Y$ are real, we impose the reality constraints
\begin{equation}
    v_m^r = \overline{v_{-m}^r}, \quad v_m^\theta = \overline{v_{-m}^\theta},\quad q_m =\overline{q_{-m}},\quad Y_m^r = \overline{Y_{-m}^r}, \quad Y_m^\theta = \overline{Y_{-m}^\theta}.
\end{equation}
Thus, we only need to consider modes for $m \geq 0$.

Since the interface is a circle, the problem is decoupled for different modes.
The separate modes $m$ create a dynamical system
\begin{equation}
    \frac{d}{dt}
    \begin{bmatrix}
        Y_m^r\\
        Y_m^\theta
    \end{bmatrix}
    =  
    \bm M_m 
    \begin{bmatrix}
        Y_m^r\\
        Y_m^\theta
    \end{bmatrix},
\end{equation}
\begin{itemize}
    \item $m=0$: $\bm{M}_0$ is a zero matrix, so $Y_0^r, Y_0^\theta$ are constants.
    \item $m=1$:
    \begin{equation}
    \bm M_1 = 
    \begin{bmatrix}
    -\frac{K_A\chi^2}{8} & -i\frac{K_A\chi^2}{8}\\
    i\frac{K_A\chi^2}{8} & -\frac{K_A\chi^2}{8}
    \end{bmatrix}.
    \end{equation}
    The two eigenpairs of ${\bm{M}_1}$ are
\begin{equation}
    \left\{0, 
    \begin{bmatrix}
        1\\
        i
    \end{bmatrix} \right\}
    , \quad\left\{-\frac{K_A\chi^2}{4},
    \begin{bmatrix}
        1\\
        -i
    \end{bmatrix}\right\}
\end{equation}
Therefore, the eigenpairs for the mode $m=1$ are
\begin{equation}
        \left\{0, 
    \begin{bmatrix}
        1\\
        0
    \end{bmatrix},
    \begin{bmatrix}
        0\\
        1
    \end{bmatrix}
    \right\}
    , \quad\left\{-\frac{K_A\chi^2}{4},
    \begin{bmatrix}
        \cos 2\theta\\
        \sin 2\theta
    \end{bmatrix},
    \begin{bmatrix}
        -\sin 2\theta\\
        \cos 2\theta
    \end{bmatrix}
    \right\}
\end{equation}
\item $m\geq 2$:
\begin{equation}
\bm M_m = 
\begin{bmatrix}
-\frac{m \paren{2m^2+K_A\paren{1-\chi^2}+\paren{\meancurv_0^2-3}}}{8} & 0\\
-i\frac{2m^2+K_A\paren{1-3\chi^2}+\paren{\meancurv_0^2-3}   }{8} & -\frac{m {K_A\chi^2}}{4}
\end{bmatrix}.
\end{equation}
The two eigenvalues of ${\bm{M}}_m$ are
\begin{equation}
    \nu_1=-\frac{{K_A\chi^2}}{4} m,\quad\nu_2=-\frac{m}{8} \paren{2 m^2+K_A\paren{1-\chi^2}+\paren{\meancurv_0^2-3}}
\end{equation}
and the associated eigenvectors are
\begin{equation}
    \bm{v}_1=
    \begin{bmatrix}
        0\\
        1
    \end{bmatrix}, \quad \bm{v}_2=
    \begin{bmatrix}
        m\\
        i
    \end{bmatrix}
\end{equation}
respectively.
Therefore, the eigenpairs for the mode $m\geq 2$ are
\begin{equation}
\begin{aligned}
        &\left\{\nu_1; \cos (m\theta) \ebtheta, \sin (m\theta) \ebtheta     \right\}, \\
    &\left\{\nu_2;  m\cos(m\theta) \ebr-\sin (m\theta) \ebtheta, m\sin(m\theta) \ebr+\cos (m\theta) \ebtheta   \right\}.
\end{aligned}
\end{equation}
\end{itemize}

According to the eigenvalues, the stability of each mode is determined as: static for $m=0$, stable for $m=1$ and for $m\ge2$, unstable when $K_A\paren{\chi^2-1}>\paren{2 m^2-3}+{\meancurv_0^2}$.
% \begin{itemize}
%     \item $m=0$: Static.
%     \item $m=1$: Stable.
%     \item $m\geq 2$: Unstable when $K_A\paren{\chi^2-1}>\paren{2 m^2-3}+{\meancurv_0^2}$.
% \end{itemize}

\bibliographystyle{apsrev4-2}
\bibliography{refs}

\end{document}